# AI, Metacognition, and the Verification Bottleneck: A Three-Wave Longitudinal Study of Human Problem-Solving

Matthias Hümmer 1*; Franziska Durner 2*; Theophile Shyiramunda 3*; Michelle J. Cummings-Koether 4*

**1* Prof. Dr.-Ing. Matthias Hümmer,**

Professor, Institute for the Transformation of Society (I-ETOS), Deggendorf Institute of Technology - European Campus Rottal-Inn, Germany

E-Mail: matthias.huemmer@th-deg.de

ORCID ID: https://orcid.org/0009-0003-8122-470X

**2* Franziska Durner, MA,**

Research Associate, Institute for the Transformation of Society (I-ETOS), Deggendorf Institute of Technology- European Campus Rottal-Inn, Germany

E-Mail: franziska.durner@th-deg.de

ORCID ID: https://orcid.org/0000-0002-8064-5364

**3* Dr. Theophile Shyiramunda,**

Postdoctoral Research Associate, Institute for the Transformation of Society (I-ETOS), Deggendorf Institute of Technology- European Campus Rottal-Inn, Germany

E-Mail: theophile.shyiramunda@th-deg.de

ORCID ID: https://orcid.org/0000-0001-6725-3756

**4* Prof. Dr. Michelle J. Cummings-Koether,**

Professor, Institute for the Transformation of Society (I-ETOS), Deggendorf Institute of Technology - European Campus Rottal-Inn, Germany

E-Mail: michelle.cummings-koether@th-deg.de

ORCID ID: https://orcid.org/0000-0002-7137-3539

**Corresponding author @ Prof. Dr.-Ing. Matthias Hümmer,**

Postdoctoral Research Associate, Institute for the Transformation of Society (I-ETOS), Deggendorf Institute of Technology- European Campus Rottal-Inn, Germany

E-Mail: matthias.huemmer@th-deg.de

ORCID ID: https://orcid.org/0009-0003-8122-470X

## Declarations

### Conflict of Interest

The authors have no relevant financial or non-financial interests to disclose.

### Funding

This research received no specific grant from any funding agency, commercial, or not-for-profit sectors.

### Authorship and Contributions

Michelle J. Cummings-Koether, and Matthias Huemmer led the conceptualization of the pilot study. The methodology for this report was developed by Matthias Huemmer and Theophile Shyiramunda. Data collection was carried out by Michelle J. Cummings-Koether, Matthias Huemmer and Franziska Durner, while formal analysis was performed by Matthias Huemmer and Theophile Shyiramunda. The original draft of the manuscript was prepared by Matthias Huemmer, with subsequent review and editing provided by Theophile Shyiramunda and Michelle J. Cummings-Koether and Matthias Huemmer. Supervision of the project was managed by Michelle Cummings-Koether and Matthias Huemmer, and project administration was undertaken by Matthias Huemmer, Michelle Cummings-Koether and Theophile Shyiramunda. All authors have read and approved the final manuscript.

### Use of Large Language Models (LLMs)

Use of AI assistance: An AI assistant (ChatGPT/GPT-5 Thinking) was used for language editing, outline refinement, and stylistic suggestions. All conceptual content, framework design, interpretations, and final decisions are the author's own. The AI system did not have authorship or decision-making roles and is not listed as an author.

### Ethics Approval and Consent to Participate

Formal ethical approval was not required for this study, as it involved minimal risk and was conducted in an institution without a dedicated ethics review board. All participants were informed about the purpose of the study and provided informed consent prior to participation. The study was conducted in accordance with recognized ethical standards for research involving human participants and applicable data protection regulations (including GDPR).

### Data Availability

All data and materials are available upon request.

Matthias Hümmer; Franziska Durner; Theophile Shyiramunda; Michelle J. Cummings-Koether

AI, Metacognition, and the Verification Bottleneck: A Three-Wave Longitudinal Study of Human Problem-Solving

**Abstract**

This longitudinal pilot study examines how generative artificial intelligence (AI) reshapes human problem-solving practices and metacognitive competencies over six months across three waves (Wave 1: n = 21; Wave 2: n = 36; Wave 3: n = 23) in an academic setting. Using mixed methods combining self-reported methodologies with objective task performance measures, we track the following dimensions: (a) problem-solving workflow evolution, (b) AI adoption patterns by task complexity, (c) consultation behavior and actual usage rates, and (d) confidence calibration with objective accuracy outcomes. By Wave 3, AI integration reached saturation with 95.7 percent daily use (compared to 52.4% in Wave 1) and 100 percent ChatGPT adoption (85.7% in Wave 1), with a dominant hybrid workflow emerging called "Think, Internet, ChatGPT, Further processing," adopted by 39.1 percent of participants (14.3% in Wave 1), representing a 2.7-fold increase in consolidated hybrid workflows over six months. Critically, we document a verification paradox: participants showed highest AI reliance for difficult tasks (73.9 percent) yet demonstrated declining verification confidence (68.1 percent in Wave 3) precisely where performance metrics showed greatest vulnerability (81 percent accuracy on difficult problems; 47.8 percent on complex tasks). Objective performance analysis revealed a clear complexity gradient with systematic accuracy decline from 95.2 percent to 81.0 percent to 66.7 percent to 47.8 percent across problem categories, accompanied by widening belief-performance gaps (plus 34.6 percentage points) and proof-belief gaps (minus 13.8 percentage points) on the most complex problems. These findings indicate a fundamental shift in the bottleneck constraining human-AI problem-solving: from solution generation to solution verification. We synthesize empirical patterns into the ACTIVE framework, which is grounded in cognitive load theory, distributed cognition, and metacognitive scaffolding, emphasizing six dimensions: Awareness and task-AI alignment assessment, Critical verification protocols with structured validation procedures, Transparent integration with human-in-the-loop governance, Iterative skill development to counter cognitive offloading, Verification confidence calibration through trust monitoring, and Ethical and contextual evaluation. We provide concrete implementation pathways for educational institutions, professional organizations, and individual practitioners to foster AI-augmented rather than AI-replacing problem-solving. Critical limitations include sample homogeneity with an academically affiliated cohort only (n=21-36 per wave, convenience sampling), potentially non-representative of corporate, clinical, legal, or other regulated professional contexts with higher verification stakes. Additional limitations include self-report bias in confidence measures (verified against objective performance, revealing 32.2 percentage point divergence), lack of control conditions precluding causal inference, restriction to mathematical/analytical problem-solving domains, and a six-month observation window insufficient to determine long-term skill trajectories or reversal patterns. Results represent exploratory rather than confirmatory findings and generalize most confidently to early-adopter, high-literacy, academically affiliated populations. Causal claims require experimental validation through randomized controlled trials.

**Keywords:** artificial intelligence, generative AI, ChatGPT, problem-solving, metacognition, cognitive offloading, verification, human-AI collaboration, hybrid intelligence, verification gap, trust calibration, AI literacy, skill preservation, longitudinal study, digital transformation

Matthias Hümmer; Franziska Durner; Theophile Shyiramunda; Michelle J. Cummings-Koether

# AI, Metacognition, and the Verification Bottleneck: A Three-Wave Longitudinal Study of Human Problem-Solving

## 1. Introduction

Generative artificial intelligence systems, particularly large language models, have rapidly transitioned from research novelties to infrastructure in knowledge work, education, and everyday problem-solving (Brynjolfsson & McAfee, 2017; Dwivedi et al., 2023). While benefits are well-documented (considering accelerated information access, rapid prototyping, code generation) fundamental questions remain about how sustained AI use transforms human problem-solving competence. Specifically: Which tasks users delegate, how workflows restructure, how users calibrate trust in AI systems, and critically, whether foundational problem-solving skills are preserved or systematically eroded through cognitive offloading. This longitudinal study addresses these questions through six months of intensive observation.

Prior work highlights two countervailing forces. On one hand, AI can increase productivity and lower barriers to complex tasks, often benefiting less-experienced users most (Li et al., 2024; Peng et al., 2023; Zhang et al. 2025). On the other, cognitive offloading - delegating memory or reasoning to tools - can dampen deliberate practice and weaken verification habits if left unmanaged (Risko & Gilbert, 2016). Human-computer interaction studies further warn that explanation fluency can inflate user confidence even when outputs are wrong, creating miscalibrated trust (Kosch et al., 2023). In educational settings, LLMs introduce heightened metacognitive demands: users must decompose tasks, interrogate outputs, track assumptions, and distinguish plausible rhetoric from warranted claims (Tankelevitch et al., 2024).

Cross-sectional snapshots cannot capture how these dynamics unfold as people acculturate to AI. Longitudinal evidence is needed to observe whether initial enthusiasm settles into sustainable hybrid workflows or drifts toward over-reliance and skill atrophy. This study addresses that gap with three survey waves over roughly six months within an academic institute. We combine broad self-reports (methodologies, application areas, frequency) with task-specific vignettes that elicit perceived complexity, consultation intent, actual use of ChatGPT, and confidence in both solution correctness and verifiability.

Three empirical patterns motivate the contribution of this paper. First, by Wave 3, AI became deeply embedded: universal ChatGPT use (100%), daily use by 95.7%, and a standardized hybrid workflow - "Think → Internet → ChatGPT → Further processing" - adopted by 39.1%. Second, AI consultation rose most for difficult tasks (to 74%), exceeded both simple (57%) and complex (48%) problems, indicating that users strategically deploy AI on well-defined but cognitively demanding work. Third, a verification gap emerged: while confidence that solutions were correct remained high, confidence in the ability to prove correctness declined for difficult and complex tasks; objective accuracy on vignette items fell with complexity (95.2% → 81.0% → 66.7% → 47.8%). Together these findings suggest a shift from generation bottlenecks to verification bottlenecks in human-AI collaboration.

We make three contributions:

1. Descriptive longitudinal evidence of cultural and methodological change in AI use, documenting the consolidation of a dominant hybrid workflow and divergence between solution confidence and verification confidence.

2. Calibration analysis linking perceived difficulty, consultation behavior, and objective accuracy, thereby clarifying where belief-performance and proof-belief gaps open as complexity increases.

3. A practical framework (ACTIVE) for sustainable integration that centers verification, metacognition, and human-in-the-loop governance to mitigate deskilling risks while preserving efficiency gains.

The remainder of the paper proceeds as follows. Section 2 outlines theoretical bases spanning cognitive offloading, trust calibration, and distributed cognition. Section 3 details the longitudinal design, instruments, and vignette tasks. Section 4 reports the results across waves, including methodology shifts, application domains, consultation patterns, confidence measures, and objective performance. Section 5 discusses implications for human-AI collaboration, verification scaffolding, and educational design, culminating in the ACTIVE framework. Section 6 operationalizes ACTIVE for institutions, workplaces, and individuals. Section 7 notes limitations and future research directions; Section 8 concludes.

## 2. Theoretical Framework

This study integrates five complementary theoretical perspectives to explain how sustained exposure to generative AI shapes human problem-solving practices.

Distributed cognition conceptualizes problem-solving as system-level processes wherein cognitive work distributes across people, tools, and representations (Hutchins, 1995). LLMs function as external representations supporting search, reformulation, and synthesis. The extended mind thesis further argues that reliably coupled tools become functional cognitive components when integrated into routines (Clark & Chalmers, 1998). This predicts consolidation of stable hybrid workflows embedding AI after initial framing but before final evaluation.

Cognitive load theory holds that working memory capacity limits performance; tools reducing extraneous load can improve outcomes (Sweller, 1988). Generative AI reduces load by supplying solutions and explanations. However, cognitive offloading research shows individuals shift computation to external media, sometimes at the expense of long-term retention (Risko & Gilbert, 2016). This creates a trade-off: offloading frees resources for higher-level reasoning but can undermine skill consolidation if it displaces essential practice.

Trust in automation effectiveness depends on appropriately calibrated reliance: over-trust produces complacency, under-trust produces disuse (Lee & See, 2004). In knowledge work, trust evolves with task experience and observed reliability (Glikson & Woolley, 2020). LLMs introduce distinctive challenges because fluent explanations inflate perceived reliability even when accuracy is uncertain (Kosch et al., 2023).

Metacognition involves monitoring and control of cognitive processes, including confidence calibration and strategy selection (Koriat, 2012). Generative AI heightens metacognitive demands requiring users to interrogate prompts, assess plausibility, and verify solutions. Without explicit scaffolds, explanation fluency produces illusions of understanding (Rozenblit & Keil, 2002).

Technology acceptance models predict utilization based on perceived usefulness, ease of use, and social norms (Venkatesh & Davis, 2000). Task-technology fit theory predicts greatest performance gains when tool capabilities align with task demands (Goodhue & Thompson, 1995). Generative AI aligns strongly with information search, summarization, and code drafting but may undermine tasks requiring formal proof or deep transfer.

Integrating these perspectives, we propose four propositions:

P1 (Hybridization): Repeated AI use yields stable hybrid workflows embedding AI after human framing and before final evaluation.

P2 (Selective Delegation): AI consultation is highest for difficult, well-structured tasks with favorable task-technology fit, lower for simple tasks and complex tasks requiring contextual judgment.

P3 (Verification Bottleneck): As AI reliance increases, solution confidence remains high while verification confidence and objective accuracy decline with complexity unless explicit verification scaffolds are employed.

P4 (Trust Calibration): Users exhibit miscalibration manifested as belief- performance and proof-belief gaps, especially for fluent but weakly grounded outputs, improving through metacognitive monitoring and triangulation. These propositions guide interpretation of results and motivate ACTIVE framework design targets: preserving human framing, enforcing verification checkpoints, supporting metacognitive monitoring, and aligning autonomy with task risk.

## 3. Materials and Methodology

### 3.1. Methodology

In its third wave (Wave 3), the longitudinal survey continues the investigation of AI, following a framework based on Huemmer et al. (2025a) methodological paper, grounded in the UTAUT model (Venkatesh et al., 2012), ensuring transparency, reproducibility, and consistency, and aligning with established survey research standards (Plano Clark, 2017) (see Figure 2). The overarching research design, including the theory-grounded framework, operationalization of constructs, and longitudinal procedure, is detailed in the companion methodological publication (Huemmer et al., 2025a). This framework prioritizes procedural transparency and reproducible reporting in line with established survey research guidelines (Plano Clark, 2017). The constructs, particularly those measuring AI adoption and perceptions, are contextualized within established technology-acceptance theories, such as the Unified Theory of Acceptance and Use of Technology (UTAUT) (Venkatesh et al., 2012), to enhance interpretability and scientific validity. Results from Wave 1 and Wave 2 of this longitudinal study, focusing on AI's intersection with culture and problem-solving, are published in Huemmer et al. (2025b), Huemmer et al. (2025c), and Cummings-Koether et al. (2025).

For Wave 3 a convenience sample of 23 participants was recruited from the European Campus Rottal-Inn community. All participants provided informed consent, and the ethical and data privacy procedures adhered to GDPR principles, remaining consistent with the protocol described in Huemmer et al. (2025a). The survey instrument consisted of structured questions grouped into key domains as outlined in Huemmer et al. (2025a) and applied consistently in Waves 1 and 2 as well as in this study (Huemmer et al., 2025b, c). These domains included demographics and role context, AI knowledge and usage patterns, perceived efficiency, reliability, safety, and ethics, problem-solving with AI, self-assessment of problem-solving methodology, and process sequences and vignettes. It should be noted that other aspects concerning the influence on human culture will be covered in a separate article and are not considered here.

Data processing and analysis were descriptive, focusing on frequency counts, percentages, and measures of central tendency (means, medians) and dispersion (standard deviations, interquartile ranges) for the entire sample. As a pilot study, no inferential tests were performed, and subgroup patterns are discussed qualitatively. The data handling procedures, from processing to reporting, followed the same reproducible procedures outlined in Huemmer et al. (2025a) and specifically as applied in Waves 1 and 2, which are also applied here for Wave 3 in Huemmer et al. (2025b, c).

Figure **1** represents the overview of the methodology used for Study Wave.

**Figure 1.** Overview of Materials and Methods for Study Wave 3.

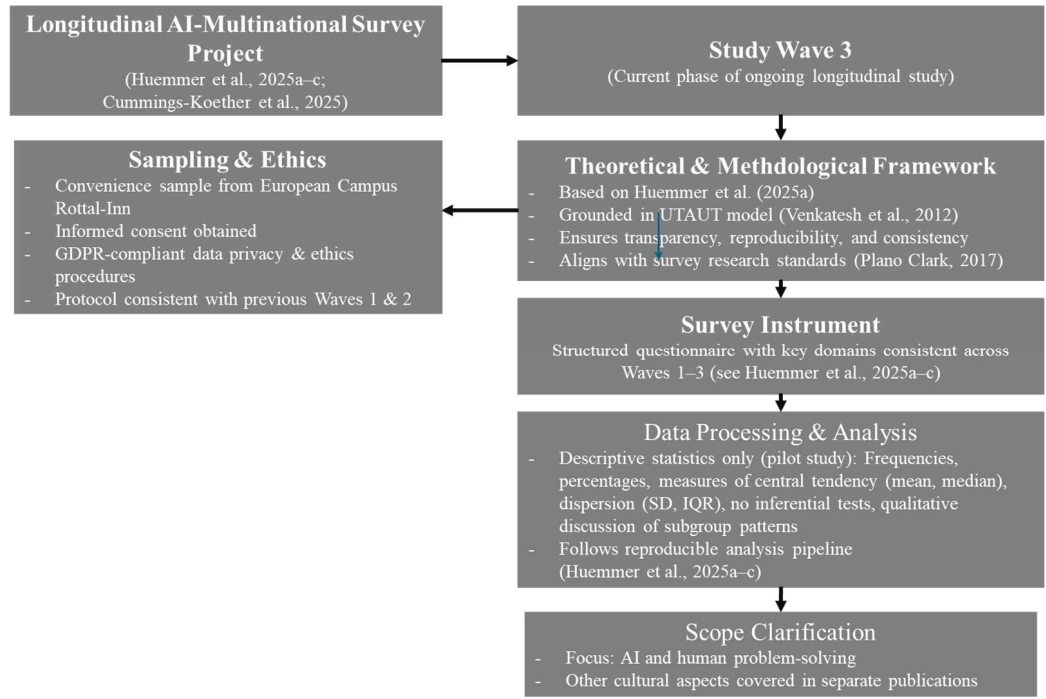

**Source.** Authors' own illustration based on the longitudinal survey framework (Huemmer et al., 2025a-c).

Figure 1 summarizes the methodological structure of Study Wave 3 within the longitudinal AI-Culture Survey Project. Building on the established framework of earlier waves (Huemmer et al., 2025a-c) and grounded in the UTAUT model, this phase integrates ethical and GDPR-compliant sampling, a structured survey instrument, and descriptive analytic procedures emphasizing transparency and reproducibility. The focus of this wave is exploratory, examining AI's role in human problem-solving, with broader cultural aspects discussed in companion studies.

Data processing and analysis were descriptive, focusing on frequency counts, percentages, and measures of central tendency (means, medians) and dispersion (standard deviations, interquartile ranges) for the entire sample. As a pilot study, no inferential tests were performed, and subgroup patterns are discussed qualitatively. The data handling procedures, from processing to reporting, followed the same reproducible procedures outlined in Huemmer et al. (2025a) and specifically as applied in Waves 1 and 2, which are also applied here for Wave 3 in Huemmer et al. (2025b, c).

## 3.2. Research design

We implemented a three-wave longitudinal panel (T1-T3) with approximately 2-3 months between waves. The study ran as a pilot inside the i-ETOS environment at Deggendorf Institute of Technology (ECRI), which enabled repeated observation of the same population and close monitoring of behavioral and cultural shifts in AI use. The complete protocol - survey instrument, item wording, and the four problem vignettes - is documented in the dedicated methodology report (Hümmer et al., 2025).

### 3.3. Participants

Participants were students and academic staff recruited via institute mailing lists, classes, and lab activities. Because panel participation was voluntary, wave-level sample sizes varied due to attrition and new entrants: Wave 1 (n = 21), Wave 2 (n = 36), Wave 3 (n = 23). All participants provided informed consent. Given the academic setting, familiarity with AI tools was common, which is typical and appropriate for the study focus. However, this academic context represents a significant boundary condition limiting generalizability. The sample's institutional affiliation, higher education status, and convenience recruitment method likely yielded participants with above-average AI literacy, intrinsic motivation for learning, and lower performance stakes compared to corporate workers in regulated industries, clinical professionals, legal practitioners, or engineers where verification failures carry substantial consequences. Results should be interpreted primarily as applying to early-adopter, high-literacy populations; findings may not replicate in occupational contexts with stricter accountability requirements, time constraints, or higher error consequences. Future research in diverse occupational contexts is essential before generalizing these findings to broader professional populations.

### 3.4. Materials and procedure

At each wave, participants completed a structured, self-administered online questionnaire comprising four components:

1. General problem-solving methodology

   Respondents indicated which methods they use (e.g., books, internet, asking others, mobile phone, AI) and could select multiple methods. Percentages reported in the results reflect the proportion of respondents endorsing each method per wave.

2. AI application areas

   Respondents reported domains in which they apply AI (e.g., ideation, research, programming, formulations, checking correctness, regular work, sketching/painting, other). Multiple selections were allowed.

3. Problem-solving with AI by perceived complexity

   Respondents indicated whether they use AI for simple, difficult, and complex problems, plus a "does not apply" option. This section captures strategic allocation of AI across task classes rather than specific tasks.

4. Task-level vignettes

   Each wave included four mathematically oriented problems representing an increasing complexity gradient (simple percentage; geometry/right triangle; combinatorics/probability; linear optimization with constraints). For each problem, respondents (a) rated perceived difficulty (simple, difficult, complex, not solvable), (b) indicated whether they would consult ChatGPT, and (c) if they used ChatGPT, rated whether they thought the solution was correct and whether they could prove its correctness. Objective correctness was coded from respondents submitted answers. Several attitude items - such as perceived reliability and data safety - used 8-point Likert scales, ensuring sensitivity to moderate shifts across waves.

Administration details

- The survey was delivered in English, hosted on a secure institutional platform.
- Participation time per wave was approximately 10-15 minutes.

- Items about tools and contexts allowed multiple responses; task-level items were single-selection except confidence/verification ratings.
- Some items were optional; therefore, per-item Ns are reported alongside percentages/means in the results for transparency.

### 3.5. Data analysis

Analyses aggregated the three waves to identify temporal patterns. Given the pilot scope and modest sample sizes, the primary approach was descriptive: counts, valid- percent percentages, means, standard deviations, medians, and interquartile ranges. For optional items, denominators reflect the number of valid responses to that item at that wave; listwise deletion was avoided to preserve information. Task-level performance was summarized as the proportion correct by problem based on submitted solutions coded as correct or incorrect by study authors. Where relevant, we contrasted consultation rates, perceived correctness, and self-assessed verifiability to illuminate belief-performance and proof-belief gaps.

A critical limitation is that confidence and verification ability measures rely entirely on self-report; no independent assessment of verification competence was conducted. Objective verification capability could be measured in future research through evaluating participants' ability to identify errors in deliberately erroneous AI solutions, assess output quality against stated criteria, or perform independent verification tasks. Visualizations and tables were generated from the same CSV exports used for computation to maintain traceability from raw responses to reported figures.

## 4. Longitudinal Results

This section presents the longitudinal findings from three waves of data collection conducted over a six-month period. The analysis tracks the evolution of AI integration into problem-solving practices, examining shifts in methodology, application domains, problem complexity preferences, and confidence in AI-generated solutions. Each subsection focuses on a specific dimension of AI adoption, revealing patterns of behavioral and cultural change as participants gained experience with AI tools.

### 4.1. Participant Demographics over the different Study Waves

The multinational longitudinal pilot study, encompassing Waves 1, 2, and 3, was designed to track the evolving dynamics of human-AI collaboration. The demographic profile of the participant sample across these waves reflects both the consistency of the sampling strategy and its evolving reach, providing important context for interpreting the longitudinal findings on AI integration and its cognitive impacts (Hümmer et al., 2025a, 2025b, 2025c).

**Table 1.** Participant Demographics Across Study Waves 1, 2, and 3.

| Characteristic | Category | Wave 1 (f) | Wave 1 (%) | Wave 2 (f) | Wave 2 (%) | Wave 3 (f) | Wave 3 (%) |
|---|---|---|---|---|---|---|---|
| **Age** | Youth (<30 years) | 9 | 42.9 | 19 | 52.8 | 12 | 52.2 |
| | Adults (30–69 years) | 12 | 57.1 | 17 | 47.2 | 11 | 47.8 |
| | Older adults (≥70 years) | 0 | 0.0 | 0 | 0.0 | 0 | 0.0 |
| **Gender** | Men | 11 | 52.4 | 23 | 63.9 | 9 | 39.1 |

| Characteristic | Category | Wave 1 (f) | Wave 1 (%) | Wave 2 (f) | Wave 2 (%) | Wave 3 (f) | Wave 3 (%) |
|---|---|---|---|---|---|---|---|
| | Women | 10 | 47.6 | 12 | 33.3 | 14 | 60.9 |
| | Diverse | 0 | 0.0 | 0 | 0.0 | 0 | 0.0 |
| | No Answer | 0 | 0.0 | 1 | 2.8 | 0 | 0.0 |
| Nationality | German | 8 | 38.1 | 10 | 28.0 | 6 | 26.1 |
| | Africa | 5 | 23.8 | 8 | 22.0 | 3 | 13.0 |
| | Asia | - | - | 6 | 17.0 | 6 | 26.1 |
| | Europe | 3 | 14.3 | 3 | 8.0* | - | - |
| | Americas | 1 | 4.8 | 3 | 8.0* | 5 | 21.7 |
| | No Answer | 4 | 19.0 | 6 | 17.0 | 3 | 13.0 |
| Occupation | Non-students | 12 | 57.1 | 14 | 38.9 | 9 | 39.1 |
| | Students | 9 | 42.8 | 22 | 61.1 | 14 | 60.9 |
| Education Level | Non-postgraduate | 7 | 33.3 | 19 | 52.8 | 11 | 47.8 |
| | Postgraduate | 13 | 61.9 | 17 | 47.2 | 12 | 52.2.8 |
| | Other | 1 | 4.8 | 0 | 0.0 | 0 | 0.0 |
| **Total Sample** | | **21** | - | **36** | - | **23** | - |

**Note.** Percentages may not sum to 100% due to rounding. Nationality categories consolidated for clarity; African and Asian responses (n<5 per wave) grouped into "Other" for confidentiality. See text (Section 4.1) for discussion of evolving demographic composition across waves.

*) Study 2 combines Europe, North and South America into single category (6 total); Study 1 reports 3 "Other" European nationalities separately

**Source.** Authors' own illustration based on survey data published in (Huemmer et al. 2025b, c, d).

As illustrated in Table 1 the study maintained a stable focus on a technologically engaged population across all waves, primarily composed of Youth and Adults, with no participants from the oldest age cohort. This consistency ensures that the observed trends in AI adoption are not confounded by significant generational digital divides. However, notable shifts are evident in other demographics. The gender balance fluctuated, with Wave 3 showing a higher proportion of women (60.9%) compared to Wave 2 (33.3%), which may influence the generalizability of specific wave-level findings related to trust and usage patterns.

A key strength of the longitudinal design is the sustained multinational diversity of the sample. While the proportion of German participants decreased from Wave 1 (38.1%) to subsequent waves (25.0% in W2, 26.1% in W3), the representation of participants from African, Asian, and American nationalities remained substantial. This evolving composition enriches the study's insights into the cultural dimensions of AI adoption first explored in Wave 1 (Cummings-Koether et al., 2025). Furthermore, the sample became increasingly student-dominated from Wave 1 (42.9%) to Waves 2 and 3 (over 60%), accompanied by a slight decrease in the proportion of postgraduate holders. This shift suggests a broader capture of perspectives from digital natives who are in the process of developing their professional problem-solving competencies.

In conclusion, the demographic trajectory across the three waves demonstrates a purposeful and dynamic sampling approach that aligns with the study's aim to investigate AI's impact within a global, academically affiliated context. The stability in age and the increased multinational and student representation strengthen the investigation into the formation of a hybrid problem-solving culture among populations who are likely to be early adopters and shapers of human-AI collaboration norms. These demographic characteristics provide a crucial lens through which the core findings on orchestrated integration and the verification gaps must be interpreted.

### 4.2. Evolution of AI Knowledge and Usage Patterns

The longitudinal design of this research program has enabled the tracking of rapid developments in AI adoption and integration across Waves 1, 2, and 3, (Hümmer et al., 2025a, 2025b, 2025c). The data reveals a clear trajectory toward near-universal familiarity and increasingly sophisticated, daily use of AI tools within the sample population, providing crucial context for the observed changes in problem-solving behaviors.

**Table 2.** Participant Demographics Across Study Waves 1, 2, and 3.

| Characteristic | Category | Wave 1 (f) | Wave 1 (%) | Wave 2 (f) | Wave 2 (%) | Wave 3 (f) | Wave 3 (%) |
|---|---|---|---|---|---|---|---|
| **AI Knowledge** | Prior knowledge in AI | 17 | 81.0 | 33 | 91.7 | 22 | 95.7 |
| | No prior knowledge | 4 | 19.0 | 3 | 8.3 | 1 | 4.3 |
| **AI Tools Used** | ChatGPT | 18 | 85.7 | 34 | 94.4 | 23 | 100.0 |
| | Claude | 4 | 19.0 | 8 | 22.2 | 5 | 21.7 |
| | Microsoft Copilot | 8 | 38.1 | 17 | 47.2 | 6 | 26.1 |
| | Google Gemini | 5 | 23.8 | 14 | 38.9 | 8 | 34.8 |
| | MetaAI | 2 | 9.5 | 10 | 27.8 | 4 | 17.4 |
| | Perplexity | 2 | 9.5 | 7 | 22.2 | 7 | 30.4 |
| | Grok | 0 | 0.0 | 8 | 22.2 | 3 | 13.0 |
| | Canva | 3 | 14.3 | 12 | 33.3 | 7 | 30.4 |
| | Others | 8 | 38.1 | 12 | 33.3 | 7 | 30.4 |
| | Private | 16 | 76.2 | 34 | 94.4 | 18 | 78.3 |

| Characteristic | Category | Wave 1 (f) | Wave 1 (%) | Wave 2 (f) | Wave 2 (%) | Wave 3 (f) | Wave 3 (%) |
|---|---|---|---|---|---|---|---|
| **Contexts of Use** | Professional | 16 | 76.2 | 23 | 63.9 | 21 | 91.3 |
| | Educational | 15 | 71.4 | 34 | 94.4 | 19 | 82.6 |
| **Frequency of Use** | Daily | 11 | 52.4 | 23 | 63.8 | 22 | 95.7 |
| | Weekly | 8 | 38.1 | 9 | 25.0 | 0 | 0.0 |
| | Other | 2 | 9.5 | 4 | 11.1 | 1 | 4.3 |
| **Total Sample** | | **21** | | **36** | | **23** | |

**Note.** Multiple responses were possible for AI tools and contexts of use. Study 1 Others: Adobe CC AI, DeepL, Deepseek, Goodnotes AI, ML tools, Elicit, Research Rabbit, TextSynth, Siri | Study 2 Others: Deepseek (5 mentions), Adobe Express, Acrobat, Elicit, ResearchRabbit, NotebookLM, Gamma, Mistral AI, Qwen, Anthropic, Deep Vision of Zeiss | Study 3 Others: Deepseek (4 mentions), Manus NotebookLM, Napkin, Ideogram, Miro, Elicit, ResearchRabbit, Gamma

**Source.** Authors' own illustration based on survey data published in (Huemmer et al. 2025b, c, d).

The data demonstrates a striking consolidation of AI knowledge and a significant intensification of use over the course of the study. Prior knowledge of AI increased substantially from Wave 1 (81.0%) to Wave 3 (95.7%), indicating the rapid saturation of AI awareness within this population. This is further corroborated by the escalating frequency of use, where daily usage more than doubled from Wave 1 (52.4%) to a near-ubiquitous 95.7% in Wave 3. This trend confirms that AI has transitioned from a novel tool to an integral, daily resource for the vast majority of participants, underpinning the observed emergence of a hybrid problem-solving culture.

A key finding across the waves is the diversification of the AI tool portfolio. While ChatGPT consolidated its position as the dominant platform, reaching 100% adoption in Wave 3, the usage of other tools evolved significantly. The entry of new, specialized tools in Wave 3, such as Perplexity (30.4%) and Canva (30.4%), alongside consistent use of Claude and Google Gemini, points to a maturing user base that is strategically selecting different AI systems for specific tasks. This tool diversification is a behavioral manifestation of the orchestrated integration observed in problem-solving sequences.

The contexts of AI use also show an important evolutionary pattern. While private and educational use remained high across all waves, professional use saw a notable surge in Wave 3 (91.3%), rebounding from a dip in Wave 2 (63.9%) to exceed the levels seen in Wave 1 (76.2%). This suggests a process of normalization and deeper integration of AI into professional workflows over time, moving beyond initial exploration phases. The consistently high application of AI for research, ideation, and checking correctness, as detailed in the core findings, is supported by this longitudinal trend of entrenchment across all life domains.

In conclusion, the longitudinal data on AI knowledge and usage patterns paints a picture of a population that is not only adopting AI rapidly but is also developing more sophisticated and differentiated engagement strategies. This evolving relationship with AI technology forms the foundational context for the study's central findings on cognitive redistribution and the critical verification gaps, explaining why these cognitive phenomena have become increasingly pronounced and measurable through each successive wave of the research.

### 4.3. Evolution of AI Perceptions Across Study Waves

The longitudinal data on participant perceptions reveals a nuanced evolution in how AI is viewed, with trust becoming more differentiated across reliability, data safety, and ethical context is summarized in Table 3. Rather than converging toward a single stable attitude, participants appear to be calibrating their trust along multiple dimensions, which helps explain the hybrid and selective problem-solving strategies observed in the study.

**Table 3.** Perceptions of AI Across Study Waves 1, 2, and 3.

| Dimension | Wave 1 % / M (SD) | Wave 2 % / M (SD) | Wave 3 % / M (SD) |
| --- | --- | --- | --- |
| **Perceived Reliability** | M = 4.1 (±1.5) | M = 5.0 (±1.2) | M = 4.9 (±1.0) |
| **Data Safety** | M = 5.9 (±2.2) | M = 5.5 (±1.7) | M = 4.9 (±1.7) |
| **Ethical Use: Private** | 76.2% "not cheating" | 83.3% "not cheating" | 65.2% "not cheating" |
| **Ethical Use: Professional** | 57.1% "not cheating" | 69.4% "not cheating" | 69.6% "not cheating" |
| **Ethical Use: Academic Theses** | 66.7% "cheating" | 38.9% "cheating" | 69.6% "cheating" |
| **Total Sample Size** | 21 | 36 | 23 |

**Note.** Reliability and safety were measured on 8-point Likert scales (1 = low/unreliable/unsafe, 8 = high/reliable/safe). Ethical use percentages reflect the proportion of respondents categorizing AI use in that context as "cheating" or "not cheating."

**Source.** Authors' own illustration based on survey data published in (Huemmer et al. 2025b, c, d).

Perceived Reliability shows a clear upward shift over time. On the 8-point scale, mean scores increase from slightly below the midpoint in Wave 1 (M = 4.1, SD = 1.5) to moderately positive levels in Waves 2 and 3 (W2: M = 5.0, SD = 1.2; W3: M = 4.9, SD = 1.0). This pattern suggests that as participants gain experience with AI tools, they come to see them as more dependable for problem-solving, without moving into uncritical enthusiasm. Reliability judgments remain in a moderate, pragmatically positive range rather than at the extremes, consistent with a view of AI as a useful but fallible collaborator.

Data Safety, in contrast, follows the opposite trajectory. Initial assessments in Wave 1 are relatively high (M = 5.9, SD = 2.2), indicating that participants start out with a comparatively strong sense that AI use is safe with respect to data handling. Over time, these ratings decline toward the scale midpoint (W2: M = 5.5, SD = 1.7; W3: M = 4.9, SD = 1.7). This downward movement suggests not a reduction of concern, but a growing caution or skepticism about how securely data are managed when using AI systems. As AI becomes more embedded in everyday practice, participants seem to develop a more critical stance on data protection and privacy issues, recalibrating their safety assessments toward a more neutral, less trusting position.

Ethical Use perceptions show the strongest context sensitivity. In private contexts, a clear majority across all waves considers AI use "not cheating," but this acceptance is not strictly stable. It rises from 76.2% in Wave 1 to 83.3% in Wave 2, then drops to 65.2% in Wave 3. This suggests that while private use remains broadly legitimate in the eyes of most participants, there is a small but notable re-tightening of norms in the final wave. Professional contexts show a more straightforward normalization pattern: acceptance of AI as "not cheating" increases from 57.1% in Wave 1 to roughly 70% in Waves 2 and 3 (69.4% and 69.6%, respectively). Here, AI appears to consolidate its status as a legitimate tool of professional practice over time.

The most pronounced shifts occur in evaluations of AI use for academic theses. The proportion of participants who classify such use as "cheating" starts high in Wave 1 (66.7%), drops markedly in Wave 2 (38.9%), and then returns to an even slightly higher level in Wave 3 (69.6%). This U-shaped pattern suggests a temporary phase of experimentation or normative uncertainty in Wave 2, followed by a reassertion of stricter academic integrity standards in Wave 3. Participants seem willing to accept AI as a normal aid in everyday and professional tasks, while simultaneously reaffirming that its use in high-stakes academic credentialing crosses an ethical boundary.

Taken together, the longitudinal perception data do not reflect a simple stabilization of trust but rather a differentiated, context-sensitive calibration. Participants become more confident in AI's reliability for solving tasks, more cautious about data safety, and more precise in drawing ethical boundaries between private, professional, and academic domains. This evolving pattern of "calibrated trust" provides the psychological foundation for the observed hybrid problem-solving strategies: AI is strategically integrated where it is seen as reliable and legitimate, held at arm's length where data risks are salient, and actively constrained in domains where it is judged to threaten core norms of fairness and academic integrity.

### 4.4. Anticipated Influence of AI on Work and Private Life

Tracking the anticipated influence of AI on working and private behavior across the three waves provides insight into how expectations evolve as participants gain real experience with AI tools. Rather than showing a simple linear increase, the longitudinal pattern reflects a combination of persistently high expectations for the future and a gradual, partly uneven adjustment of present-day assessments, see Table 4.

**Table 4.** Perceptions of AI Across Study Waves 1, 2, and 3.

| Behavior | Timeframe | | Wave 1 Influence Score | Wave 2 Influence Score | Wave 3 Influence Score |
|---|---|---|---|---|---|
| **Working Behavior** | Present | | M = 4.9 (±2.1) | M = 4.7 (±2.0) | M = 5.0 (±2.0) |
| | Future (2-5 years) | | M = 6.2 (±2.2) | M = 6.1 (±1.7) | M = 5.9 (±2.1) |
| **Private Behavior** | Present | | M = 3.8 (±2.0) | M = 3.6 (±2.0) | M = 4.1 (±2.3) |
| | Future (2-5 years) | | M = 5.3 (±2.1) | M = 5.3 (±2.0) | M = 5.1 (±2.1) |
| **Gap (Future - Present)** | Working Behavior | | +1.3 | +1.4 | +0.9 |
| | Private Behavior | | +1.5 | +1.7 | +1.0 |

**Source.** Authors' own illustration based on survey data published in (Huemmer et al. 2025b, c, d).

Future influence ratings are consistently high across all three waves. For working behavior, expected influence over the next two to five years remains in the upper range of the 8-point scale (Wave 1: M = 6.2, SD = 2.2; Wave 2: M = 6.1, SD = 1.7; Wave 3: M = 5.9, SD = 2.1). Private behavior follows a similar pattern at a slightly lower level (W1 and W2: M = 5.3, SD = 2.1 and 2.0; W3: M = 5.1, SD = 2.1). These values indicate a stable belief that AI will play a strong and growing role in both professional and private domains, even though there is a slight softening of the most optimistic expectations by Wave 3, particularly in the work context.

Present-day influence ratings show a more mixed but informative picture. For working behavior, current influence is already moderate and remains relatively stable across waves (W1: M = 4.9, SD = 2.1; W2: M = 4.7, SD = 2.0; W3: M = 5.0, SD = 2.0), suggesting that AI is perceived as a meaningful, but not yet dominant, factor in everyday work. For private behavior, present influence starts lower but ends slightly higher (W1: M = 3.8, SD = 2.0; W2: M = 3.6, SD = 2.0; W3: M = 4.1, SD = 2.3). The dip in Wave 2 followed by an increase in Wave 3 points to a temporary phase of cautious reassessment before participants ultimately acknowledge a somewhat stronger AI footprint in their private lives.

The gap between future and present influence scores provides a useful indicator of how far ahead participants expect AI's role to move relative to their current experience. For working behavior, this gap is +1.3 in Wave 1, increases slightly to +1.4 in Wave 2, and then contracts to +0.9 in Wave 3. For private behavior, it widens from +1.5 (Wave 1) to +1.7 (Wave 2), before narrowing to +1.0 in Wave 3. Taken together, these patterns do not show a simple, monotonic convergence, but rather an expectation "overshoot" in Wave 2 followed by a recalibration in Wave 3. By the final wave, participants still expect AI to become more influential than it currently is in both domains, yet the distance between present and future influence is noticeably smaller than at the outset, especially in private life. This suggests that some of the anticipated future impact is indeed being realized, while at the same time future expectations are being adjusted to more realistic levels.

Across all three waves, AI is consistently perceived as more influential in work than in private life, both now and in the foreseeable future. Present influence in the workplace remains higher than in private contexts in every wave, and expected future influence follows the same pattern. This aligns with the usage data showing strong professional adoption and supports the interpretation that participants primarily frame AI as productivity and performance technology, with personal use lagging somewhat behind but gradually increasing.

Overall, the longitudinal trajectory of perceived influence suggests that participants continue to view AI as a significant driver of change, particularly in their working lives, while gradually aligning their expectations with lived experience. The combination of persistently high future ratings, modest increases in present influence (especially in private contexts), and a shrinking expectation gap from Wave 1 to Wave 3 is consistent with the emergence of a more grounded, experience-based understanding of AI. This perceptual adjustment provides an important backdrop for the behavioral findings documented in this study, including the entrenchment of hybrid problem-solving practices and the strategic, orchestrated integration of AI into everyday work and, increasingly, private routines.

### 4.5. Problem-Solving Methodology and AI Integration

### 4.5.1. Applied Problem-Solving Methodologies

An analysis of the participants' problem-solving steps for difficult or complex problems shows a gradual restructuring of workflows toward hybrid, AI-supported methodologies rather than a simple replacement of traditional approaches. Table 5 documents how participants combine human thinking, analog tools, internet resources, and ChatGPT across the three waves, revealing an increasingly patterned ecosystem of human–AI co-problem-solving.

**Table 5.** Problem-Solving Workflow Patterns Across Three Waves

| Workflow Pattern | Wave 1 (n=21) | Wave 2 (n=35) | Wave 3 (n=23) |
| --- | --- | --- | --- |
| Think, Paper, Sketch, Book, Further Processing | 38.1% | 17,1% | 21.7% |
| Think, ChatGPT, Further Processing | 4.8% | 25.7% | 8.7% |
| Think, Internet, Further Processing | 9.5% | 5.7% | 0.0% |
| Think, Internet, ChatGPT, Further Processing | 14.3% | 25.7% | 39.1% |
| ChatGPT, Think, Further Processing | 4.8% | 5.7% | 4.3% |
| ChatGPT, Internet, Think, Further Processing | 0.0% | 5.7% | 13.0% |
| ChatGPT, Reformulation, ChatGPT, Reformulation | 0.0% | 0.0% | 0.0% |
| Other | 28.6% | 14.3% | 13.0% |

**Source.** Authors' own illustration based on survey data published in (Huemmer et al. 2025b, c, d).

Table 5 shows the traditional analog workflow "Think, Paper, Sketch, Book, Further Processing" remains important but clearly loses its early dominance. It is the most frequent pattern in Wave 1 (38.1 percent), drops sharply in Wave 2 (17.1 percent), and then partially recovers in Wave 3 (21.7 percent). Overall, this represents a substantial net decline of around sixteen percentage points across the study, indicating that while analog methods continue to play a role, they are no longer the default strategy for many participants.

The most prominent growth trajectory appears in the hybrid workflow "Think, Internet, ChatGPT, Further Processing." This pattern starts at 14.3 percent in Wave 1, increases to 25.7 percent in Wave 2, and reaches 39.1 percent in Wave 3, where it becomes the single most frequent workflow, used by roughly two out of five participants. The sequence places human cognition at the beginning, followed by structured consultation of internet resources and then ChatGPT before final processing. This pattern suggests a layered approach in which participants first frame the problem themselves, then draw on both conventional web search and AI-generated input, and only afterward synthesize and refine the solution.

The direct "Think, ChatGPT, Further Processing" workflow shows a more volatile pattern. It is rare in Wave 1 (4.8 percent), peaks sharply in Wave 2 (25.7 percent), and then drops again to 8.7 percent in Wave 3. This suggests that during the second wave many participants experimented with a relatively direct reliance on ChatGPT immediately after their initial thinking phase, but by the third wave a substantial share of these users appear to have migrated toward more complex hybrid workflows that also involve internet search, such as "Think, Internet, ChatGPT." What remains in Wave 3 is a smaller but stable group who continue to engage with ChatGPT directly after an initial individual reasoning phase.

The purely internet-based pattern "Think, Internet, Further Processing" gradually loses relevance and disappears completely by the end of the study. It declines from 9.5 percent in Wave 1 to 5.7 percent in Wave 2 and reaches 0 percent in Wave 3. This trajectory does not support the idea of an immediate extinction "after Wave 1," but it does indicate that by the final wave participants no longer rely on web search alone for complex problem-solving. Instead, those who previously used the internet as their main external resource appear to have transitioned to workflows that combine internet search with ChatGPT.

AI-first workflows, where ChatGPT precedes human thinking, remain a minority but show some growth and differentiation. The pattern "ChatGPT, Think, Further Processing" stays consistently low across all waves (4.8, 5.7, and 4.3 percent), indicating that only a small fraction of participants uses ChatGPT as an initial trigger before engaging in deeper individual reasoning. In contrast, the more elaborate "ChatGPT, Internet, Think, Further Processing" pattern emerges only from Wave 2 onward, increasing from 5.7 percent to 13.0 percent in Wave 3. This sequence suggests a distinct problem-solving philosophy in which participants use ChatGPT to obtain an initial orientation or problem framing, then validate or expand this with targeted internet searches, and only afterward consolidate their own thinking and solution.

The category "Other" shrinks from 28.6 percent in Wave 1 to 14.3 percent in Wave 2 and 13.0 percent in Wave 3. This reduction indicates a progressive consolidation around a limited set of recognizable workflows, particularly the hybrid patterns that combine thinking, internet, and ChatGPT. Notably, the explicitly iterative pattern "ChatGPT, Reformulation, ChatGPT, Reformulation" is not adopted at all in any wave (0 percent throughout). Participants do not shift toward purely AI-driven refinement loops; instead, the human element remains embedded as either the starting point, the central processing step, or the final integrative instance in all major workflows.

Taken together, the longitudinal data in Table 5 reveal not a replacement of human problem-solving by AI, but the emergence of a structured repertoire of hybrid workflows. Traditional analog methods are still present but significantly reduced, internet-only approaches are phased out by the final wave, and combined "Think, Internet, ChatGPT" sequences become the dominant problem-solving pattern. Across these shifts, human cognition remains the anchor of the process, with ChatGPT and internet resources orchestrated around it rather than taking over entirely. This pattern is consistent with the broader picture of calibrated, strategically integrated AI use observed throughout the study.

### 4.5.2. Longitudinal Evolution of Problem-Solving Strategies Across Complexity Levels

The longitudinal data across Waves 1, 2, and 3 shows a differentiated but coherent trajectory in how participants adapt their problem-solving methodologies to incorporate AI. Rather than a simple linear increase in AI reliance, Table 6 reflects a gradual refinement of strategy: stable self-reliance for simple problems, progressively more structured human–AI collaboration for difficult problems, and highly organized, multi-source workflows with built-in validation for complex problems. This pattern provides behavioral evidence for an emerging paradigm of orchestrated integration rather than replacement.

**Table 6.** Evolution of Problem-Solving Strategies Across Waves 1, 2, and 3.

| Problem Category | Wave 1 Strategy | Wave 2 Strategy | Wave 3 Strategy |
|---|---|---|---|
| **Simple Problems** | Direct cognitive strategy. "Think - Solve" is dominant, with AI as a secondary tool for confirmation when needed. | Direct cognitive strategy with a preference for efficiency and autonomy. "Think - Solve" is | Direct cognitive strategy with strong self-reliance. Sequences like "Think - Solve" dominate, with AI used selectively for |

| Problem Category | Wave 1 Strategy | Wave 2 Strategy | Wave 3 Strategy |
|---|---|---|---|
| | | primary, with AI for rapid confirmation. | verification only after initial attempts. |
| **Difficult Problems** | Systematic approach emphasizing decomposition. AI evolves into a brainstorming aid or resource after individual efforts are exhausted. | Systematic approach incorporating structured decomposition and prototyping. AI becomes a collaborative partner in middle stages after initial effort. | Systematic approach with deliberate integration of multiple tools (sketching, Internet, AI). AI acts as a collaborative partner for analysis and refinement after individual effort. |
| **Complex Problems** | Maximum structure and resource integration. Decomposition is paramount, with AI embedded within multi-stage workflows that include explicit validation phases. | Maximum structure with extensive resource integration and iterative validation. Use of specialized resources and multiple AI systems for comparative analysis. | Comprehensive resource integration and iterative validation. Use of diverse sources (books, papers) and multiple AI verification cycles in structured, multi-stage workflows. |

**Source.** Authors' own illustration based on survey data published in (Huemmer et al. 2025b, c, d).

For simple problems, the strategies remain remarkably stable across all three waves. In every wave, participants rely primarily on direct cognitive approaches, with "Think – Solve" as the dominant pattern and AI used only as an optional verification tool. Wave 3 strengthens the emphasis on self-reliance, describing "strong self-reliance" and selective use of AI only after initial attempts. Wave 2 explicitly highlights efficiency and autonomy, but the basic structure remains the same. This stability indicates that the presence of AI does not erode cognitive autonomy for straightforward tasks; instead, AI is consistently positioned as a quick checker rather than a primary problem solver in this domain.

For difficult problems, the evolution is more pronounced. In Wave 1, participants follow a systematic, decomposition-based approach in which AI is introduced only after individual efforts have been exhausted and function mainly as a brainstorming aid or additional resource. By Wave 2, the same class of problems is tackled with more structured decomposition and prototyping, and AI is explicitly framed as a collaborative partner in the middle stages, once a first individual structure has been established. In Wave 3, this trajectory culminates in a deliberately multi-tool strategy that integrates sketching, internet resources, and AI. Here, AI is no longer a late add-on or emergency option, but a regular collaborator used for analysis and refinement after an initial cognitive phase. This progression illustrates how participants learn to place AI in the middle of their workflow rather than at the margins.

For complex problems, high structure and resource integration are already present in Wave 1. Participants decompose the task, draw on multiple resources, and embed AI within multi-stage workflows that include explicit validation phases. In Wave 2, these workflows become even more elaborate, with "maximum structure," extensive resource integration, and the use of specialized resources and multiple AI systems for comparative analysis. Wave 3 maintains this high structural complexity but shifts the emphasis toward "comprehensive resource integration and iterative validation," involving diverse sources such as books and papers combined with multiple AI verification cycles. The role of AI in complex problem-solving therefore does not move from absence to presence; rather, it evolves from being one component in a structured workflow to being part of repeated, multi-step verification and refinement cycles.

Against this backdrop, the most important longitudinal trend is not simply "more AI," but the progressive normalization and strategic embedding of AI within differentiated workflows. In the early stages, AI appears as a secondary resource or brainstorming aid, particularly for difficult problems after individual effort. Over time, it becomes a standard element of systematic, multi-tool workflows, especially for difficult and complex problems where it is used for analysis, refinement, and iterative validation. At the same time, its role for simple problems remains intentionally constrained to optional verification, preserving a clear boundary between tasks where human cognition is sufficient and those where augmentation is perceived as beneficial.

This evolution aligns with the core cognitive findings of the study. High consultation rates for complex tasks are mirrored in the structured, multi-source workflows described in Table 6, while the emergence of iterative AI verification cycles for complex problems in later waves can be interpreted as a behavioral strategy to manage known verification challenges. Participants are not only integrating AI into their workflows, but they are also building procedural safeguards around it, distributing cognitive labor across humans, traditional resources, and AI while retaining responsibility for final validation.

In conclusion, the longitudinal trajectory of problem-solving strategies reveals a population that is learning to use AI more deliberately rather than more indiscriminately. The progression from basic decomposition to comprehensive, iterative, and multi-source workflows, especially for difficult and complex problems, reflects a maturation of human–AI collaboration practices. This behavioral evolution toward orchestrated integration provides the methodological backbone of the emerging collaboration paradigm, in which AI is embedded as a calibrated collaborator and validator, while humans retain strategic control and interpretive authority.

### 4.5.3. Means applied for Problem-Solving

Table 7 presents the percentage of participants utilizing various problem-solving methods across the three waves and allows a differentiated view of how traditional and digital approaches evolved over time.

**Table 7.** Problem-Solving Methodology Usage Across Three Waves

| Methodology | Wave 1 (n=21) | Wave 2 (n=36) | Wave 3 (n=23) |
| --- | --- | --- | --- |
| Books | 81.0% | 72.0% | 69.6% |
| Internet | 100.0% | 94.4% | 95.7% |
| Mobile phone | 66.7% | 58.3% | 60.9% |
| Library | 52.4% | 44.4% | 47.8% |
| Asking others | 71.4% | 94.4% | 78.3% |
| Guessing/Estimation | 52.4% | 55.6% | 43.5% |
| Other | 14.3% | 8.2% | 13.0% |

**Source.** Authors' own illustration based on survey data published in (Huemmer et al. 2025b, c, d).

Overall, the data in Table 7 suggest a modest rebalancing rather than a dramatic shift away from traditional methods. Book usage remains high throughout, declining from 81.0 percent in Wave 1 to 72.0 percent in Wave 2 and 69.6 percent in Wave 3. This represents a reduction of around eleven percent points over the study but still leaves books as a frequently used resource for a clear majority of participants. Library use shows a similar pattern of attenuation without collapse. It drops from 52.4 percent in Wave 1 to 44.4 percent in Wave 2, then partially recovers to 47.8 percent in Wave 3. These trajectories indicate that traditional, text-based resources continue to play an important role, even as their relative prominence is slightly reduced.

Internet use remains the dominant methodology across all three waves. It is universally used in Wave 1 (100.0 percent) and stays near universal in Waves 2 and 3 (94.4 and 95.7 percent). This consistency underlines the internet's status as the foundational problem-solving resource in this cohort. The small dip in Wave 2 followed by a marginal increase in Wave 3 is more consistent with normal variation than with a substantive behavioral shift.

Interpersonal and informal strategies show more fluctuation. "Asking others" rises sharply from 71.4 percent in Wave 1 to 94.4 percent in Wave 2, before decreasing to 78.3 percent in Wave 3. This suggests that social consultation is a robust and recurrent strategy, with Wave 2 possibly reflecting a phase of particularly intensive peer exchange, for example when new tools or practices are being negotiated. Guessing or estimation increases slightly from 52.4 percent in Wave 1 to 55.6 percent in Wave 2, then drops to 43.5 percent in Wave 3. The final value is noticeably lower than the initial one, which supports the interpretation that, over time, readily available digital and AI-supported resources reduce the need to rely on intuitive approximation as a primary method.

Mobile phone usage shows a relatively stable pattern with a slight net decrease. It is used by 66.7 percent of participants in Wave 1, falls to 58.3 percent in Wave 2, and then rises again to 60.9 percent in Wave 3. This suggests that mobile devices remain an important, but not exclusive, access channel, and that some participants may be shifting part of their problem-solving activity to laptops or other platforms better suited for intensive AI and internet use. The "Other" category is used by a minority throughout (14.3, 8.2, and 13.0 percent), with a dip in Wave 2 and a partial rebound in Wave 3, indicating some diversity and experimentation in methods beyond the main categories, but without a clear directional trend.

In sum, Table 7 does not support the idea of a wholesale replacement of traditional methods by digital or AI-mediated approaches. Instead, it points to a layered configuration in which internet-based methods are universally established, books and libraries remain widely used but slightly less central, interpersonal consultation is actively employed and guessing declines somewhat as more reliable resources become routinely available. This configuration provides the methodological backdrop for the more fine-grained hybrid workflows and orchestrated integration patterns observed in later sections.

### 4.5.4. Functional Domains and Workflow Stages of AI Deployment

The data in **Table 8** show a clear reconfiguration of how participants apply AI across task domains, with use shifting over time from exploratory and generative functions toward more targeted, utility-oriented applications such as research, verification, and regular work tasks.

**Table 8.** AI Application Areas Across Three Waves.

| Application Area | Wave 1 (n=21) | Wave 2 (n=36) | Wave 3 (n=23) |
|---|---|---|---|
| Ideation | 66.7% | 61.1% | 47.8% |

Matthias Hümmer; Franziska Durner; Theophile Shyiramunda; Michelle J. Cummings-Koether

AI, Metacognition, and the Verification Bottleneck: A Three-Wave Longitudinal Study of Human Problem-Solving

| Application Area | Wave 1 (n=21) | Wave 2 (n=36) | Wave 3 (n=23) |
| --- | --- | --- | --- |
| Creative works | 23.8% | 52.8% | 30.4% |
| Regular works | 19.0% | 50.0% | 65.2% |
| Formulations | 66.7% | 61.1% | 43.5% |
| Painting | 9.5% | 5.6% | 8.7% |
| Sketching | 4.8% | 27.8% | 13.0% |
| Programming | 28.6% | 52.8% | 34.8% |
| Checking of correctness | 52.4% | 69.4% | 69.6% |
| Research | 52.4% | 72.2% | 82.6% |
| Other | 19.0% | 0.0% | 13.0% |

**Source.** Authors' own illustration based on survey data published in (Huemmer et al. 2025b, c, d).

Research emerges as the most clearly consolidating application area. AI use for research rises steadily from 52.4 percent in Wave 1 to 72.2 percent in Wave 2 and 82.6 percent in Wave 3, making it the single most common use case by the end of the study. This continuous increase suggests that participants increasingly treat AI as a routine component of information search, synthesis, and literature-related work, embedding it into their core knowledge workflows.

Checking of correctness follows a similar, though slightly flatter, upward trajectory. It starts at 52.4 percent in Wave 1, climbs to 69.4 percent in Wave 2, and stabilizes at 69.6 percent in Wave 3. By the final wave, roughly seven out of ten participants use AI to review or validate their outputs. This stabilization points to a durable verification role for AI: not just generating content, but cross-checking calculations, formulations, or reasoning steps in a way that becomes a standard element of many workflows.

Regular works show the strongest relative expansion. Use of AI for routine or standard tasks increases from 19.0 percent in Wave 1 to 50.0 percent in Wave 2 and 65.2 percent in Wave 3. This substantial growth indicates that AI progressively moves from a specialized or experimental tool into a normal part of everyday work practices. By Wave 3, AI is no longer confined to special projects or creative experiments; it is embedded in regular operational activities for a clear majority of participants.

At the same time, several cognitively "front-loaded" uses decline. AI for ideation drops from 66.7 percent in Wave 1 to 61.1 percent in Wave 2 and 47.8 percent in Wave 3. Formulations follow a comparable pattern, moving from 66.7 percent to 61.1 percent and then to 43.5 percent. These downward trends suggest refinement in how participants apply AI during early stage thinking and writing. Rather than using AI broadly for initial ideas and phrasing, participants appear to become more selective, shifting emphasis toward downstream support functions such as research, correctness checking, and routine execution, where perceived utility may be higher and ethical or authenticity concerns lower.

Creative and technical applications show a wave-shaped pattern. AI use for creative works increases from 23.8 percent in Wave 1 to 52.8 percent in Wave 2, then decreases to 30.4 percent in Wave 3. Programming follows a similar trajectory, rising from 28.6 percent to 52.8 percent and then moderating to 34.8 percent. These profiles suggest an exploratory peak in Wave 2, where many participants test AI in creative and coding contexts, followed by a consolidation phase in Wave 3 in which some users retain these practices while others scale back, possibly after evaluating the practical benefits and limitations in their own domains.

Visual creative tasks remain comparatively niche but not static. AI use for painting stays low across all waves (9.5, 5.6, and 8.7 percent), indicating that only a small minority consistently experiment with AI-generated images or artwork. Sketching, by contrast, shows a temporary surge: it is rare in Wave 1 (4.8 percent), reaches 27.8 percent in Wave 2, and then drops to 13.0 percent in Wave 3, ending clearly above its initial level. This pattern points to short-term experimentation with AI-supported visual thinking tools, with a smaller group continuing to use them in a more sustained way.

Finally, the "Other" category decreases from 19.0 percent in Wave 1 to 0.0 percent in Wave 2 and then reappears at 13.0 percent in Wave 3. The temporary disappearance suggests that, during Wave 2, most AI uses could be captured within the main predefined categories, while the renewed presence in Wave 3 hints at emerging niche or specialized applications that fall outside the original schema.

Taken together, the longitudinal data in Table 8 do not simply show "more AI use," but a redistribution of where AI is applied. Early, broad use in ideation and formulation is gradually complemented and partly replaced by more focused deployment in research, correctness checking, and regular work tasks. Creative and technical uses peak during an exploratory middle phase and then settle at moderate levels. This pattern is consistent with a maturing user population that experiments widely, evaluates outcomes, and then anchors AI where it delivers the most stable and perceived value within their everyday workflows.

### 4.5.5. AI Usage Depending on Problem Categories

AI deployment was reported across all problem complexity classifications. Table 9 shows how often participants used AI for simple, difficult, and complex problems in each wave and how these patterns gradually differentiated over time.

**Table 9.** AI Adoption Rates Stratified by Problem Complexity Category - Across Three Waves

| Problem Complexity | Wave 1 (n=21) | Wave 2 (n=36) | Wave 3 (n=23) |
| --- | --- | --- | --- |
| Simple | 61.9% | 58.0% | 56.5% |
| Difficult | 52.4% | 52.8% | 73.9% |
| Complex | 47.6% | 44.4% | 47.8% |
| Does not apply | 14.3% | 8.3% | 4.3% |

**Source.** Authors' own illustration based on survey data published in (Huemmer et al. 2025b, c, d).

For simple problems, AI usage remains relatively stable but shows a slight downward drift. Adoption starts at 61.9 percent in Wave 1, moves to 58.0 percent in Wave 2, and reaches 56.5 percent in Wave 3. Across all waves, roughly three out of five participants use AI even for straightforward tasks. This suggests that AI is consistently viewed as helpful for efficiency or quick checking, but there is no trend toward either strong expansion or rejection in this category. Simple tasks remain an area where AI is useful but not indispensable.

The most pronounced change occurs for difficult problems. AI usage in this category is essentially flat between Wave 1 (52.4 percent) and Wave 2 (52.8 percent), then rises sharply to 73.9 percent in Wave 3. This increase of more than twenty percent in the final wave marks a clear shift in how participants handle challenging but well-structured tasks. By the end of the study, difficult problems are the category with the highest AI adoption rate, indicating that participants have come to see AI as particularly valuable where problems demand more cognitive effort but remain sufficiently defined to benefit from structured assistance.

AI usage for complex problems remains comparatively stable and consistently lower than for simple and difficult problems. The adoption rates fluctuate only slightly around the mid-range (47.6 percent in Wave 1, 44.4 percent in Wave 2, 47.8 percent in Wave 3). This pattern suggests that, despite overall normalization of AI use, participants remain cautious about relying on AI for tasks they perceive as genuinely complex, often involving ambiguity, multiple interdependencies, or high stakes. In such cases, AI is used by about half of the participants, but it does not become the dominant strategy.

The "Does not apply" category declines steadily from 14.3 percent in Wave 1 to 8.3 percent in Wave 2 and 4.3 percent in Wave 3. By the final wave, more than 95 percent of participants report using AI for at least one of the problem categories. This reduction indicates that AI is increasingly seen as relevant across the full spectrum of problem types, even if its intensity of use still varies by complexity level.

Taken together, the pattern in Table 9 reflects a strategic maturation in AI deployment. Participants maintain a stable, moderate level of AI use for simple tasks, substantially expand their use of AI for difficult problems where structured cognitive support is especially valuable, and remain more selective and cautious in applying AI to complex problems. This differentiation suggests that participants are not indiscriminately increasing AI usage; instead, they are learning to align AI involvement with the structure and demands of the task, reinforcing the broader picture of calibrated, context-sensitive human–AI integration observed throughout the study.

### 4.6. Task-Specific Problem-Solving: AI Consultation and Performance

#### 4.6.1. Problem-Solving Vignettes: Design and Rationale

As shown in Huemmer et al. (2025a, b, c, d), the survey employed four problem vignettes arranged along a graduated complexity continuum to probe how AI use varies with cognitive demand, following established sequenced-task paradigms for studying cognitive offloading (Risko & Gilbert, 2016). The same vignette set was administered in Wave 1 (Huemmer et al., 2025b) and Wave 2 (Huemmer et al., 2025c). Problem 1 (Simple) was "What is 20% of 150?". It operationalizes a baseline arithmetic task to test thresholds of "cognitive convenience" and effort-conservation usage (Gerlich, 2024, 2025), amid evidence that even elementary calculations are offloaded to digital tools (Shanmugasundaram & Tamilarasu, 2023). Problem 2 (Difficult) regarded a question about a triangle right-angle identification via the Pythagorean theorem. It allows us to assess whether AI is invoked for verification versus solution generation when declarative and procedural knowledge are required; it also probes depth-effort trade-offs (Stadler et al., 2024) and the expertise-reversal effect in human-AI collaboration (Chen et al., 2017). Problem 3 (Complex) was a two-draw dependent-probability item. It targets multi-step combinatorial reasoning that elevates intrinsic load (Sweller, 1988), thereby testing AI-dependency in chained reasoning and confidence-verification mismatches (Zhang et al., 2024; Lee et al., 2025), including risks of metacognitive laxity under assistance (Fan et al., 2024). Problem 4 (Complex/Linear Programming) utilized a constrained profit-maximization model, which captures the apex of modeling complexity where constraint satisfaction, counterintuitive optima, and prompt specificity shape outcomes; it thus interrogates known AI weaknesses in constraint-based reasoning and the need for prompt optimization (Chen et al., 2024; Shojaee et al., 2025). Sequencing was informed by Cognitive Load Theory and recent AI-assisted problem-solving research (Kasneci et al., 2023; Sweller, 1988), enabling analysis of how computational intensity, conceptual depth, stepwise integration, and modeling demands systematically modulate AI engagement, self-efficacy, and verification behavior across the problem-solving workflow.

### 4.6.2. Participant Self-Assessment of AI Output Accuracy Across Problem Complexity Level

Participants' perceptions of artificial intelligence (AI) system reliability demonstrate a complexity-dependent pattern of calibrated trust rather than uniform confidence across all problem categories. This section presents longitudinal data across three survey waves examining how participants assessed the correctness of AI-generated outputs as a function of problem difficulty.

**Overall Pattern and Calibration**

The empirical findings reveal a systematic relationship between perceived problem complexity and reported confidence in AI accuracy. Participants exhibited highest confidence in AI flawlessness when encountering simple problems and progressively lower confidence as problem complexity increased. This pattern held consistently across all three measurement waves, suggesting that participants developed and maintained appropriate calibration of their reported expectations relative to task difficulty. However, critical analysis presented in Section 4.6.7 (Longitudinal Analysis of Problem-Solving Performance and Verification Gaps) and Tables 22-23 demonstrate that reported confidence diverges substantially from objective accuracy (e.g., 93.8% reported correctness confidence with only 47.8% actual accuracy for Problem 4 in Wave 3). These confidence ratings therefore reflect participants' subjective experience rather than objective verification competence. All confidence claims in this section should be interpreted as reported confidence rather than verified or validated confidence. Importantly, the prevalence of intermediate response categories (encompassing partial or majority correctness judgments) indicates that participants engaged in nuanced assessments rather than adopting dichotomous all-or-nothing perspectives.

**Simple Problem Category**

For simple problems, participants' confidence remained relatively high throughout the study period (Table 10). The proportion of participants indicating that all results were correct increased from 22.2% in Wave 1 to 25.0% in Wave 2, then increased substantially to 47.8% in Wave 3. The modal response category across all three waves was "majority of results are correct," accounting for 61.1%, 62.5%, and 30.4% of participants in Waves 1, 2, and 3 respectively. Notably, the combined proportion of participants endorsing either complete or majority correctness ranged from 83.3% to 100% across all waves. Only a small fraction of participants (12.5%–21.7%) selected "some results are correct," and no participants reported that all results were incorrect.

**Table 10.** Participants belief about answer correctness of AI-generated results for problem category Simple

| Correctness of AI-generated results | Problem Category Simple | | |
|---|---|---|---|
| | Wave 1 | Wave 2 | Wave 3 |
| All results are correct | 22.2% | 25.0% | 47.8% |

Matthias Hümmer; Franziska Durner; Theophile Shyiramunda; Michelle J. Cummings-Koether

AI, Metacognition, and the Verification Bottleneck: A Three-Wave Longitudinal Study of Human Problem-Solving

| Correctness of AI-generated results | Problem Category Simple | | |
| --- | --- | --- | --- |
| | Wave 1 | Wave 2 | Wave 3 |
| Some results are correct | 16.7% | 12.5% | 21.7% |
| Majority of results are correct | 61.1% | 62.5% | 30.4% |
| No results are correct | 0.0% | 0.0% | 0.0% |
| No Answers | 0.0% | 0.0% | 0.0% |
| Total Answers | n=19 | n=32 | n=23 |

**Source.** Authors' own illustration based on survey data published in (Huemmer et al. 2025b,c,d).

**Difficult Problem Category**

For difficult problems, participant assessments revealed greater heterogeneity and reduced confidence relative to the simple category (Table 11). In Wave 1, the preponderance of responses (57.9%) fell into the "some results are correct" category. By Wave 2, responses shifted substantially toward the modal category of "majority of results are correct" (50.0%), with only 28.1% maintaining the "some results are correct" assessment. Wave 3 data showed a convergence between these two categories, with 36.4% selecting "majority of results are correct" and 27.3% selecting "some results are correct." The proportion indicating complete correctness increased from 15.8% in Wave 1 to 21.9% in Wave 2, before increasing further to 36.4% in Wave 3. Consistent with findings for simple problems, no participants reported complete failure of AI outputs.

**Table 11.** Participants belief about answer correctness of AI-generated results for problem category Difficult.

| Correctness of AI-generated results | Problem Category Difficult | | |
| --- | --- | --- | --- |
| | Wave 1 | Wave 2 | Wave 3 |
| All results are correct | 15.8% | 21.9% | 36.4% |
| Some results are correct | 57.9% | 28.1% | 27.3% |
| Majority of results are correct | 26.3% | 50.0% | 36.4% |
| No results are correct | 0.0% | 0.0% | 0.0% |
| No Answers | 0.0% | 0.0% | 0.0% |
| Total Answers | n=18 | n=32 | n=22 |

**Source.** Authors' own illustration based on survey data published in (Huemmer et al. 2025b,c,d).

**Complex Problem Category**

For complex problems, the distribution of participant assessments reflected substantially reduced confidence in AI output quality (Table 12). The modal response category across all waves was "some results are correct," which represented 63.2%, 48.5%, and 54.5% of responses in Waves 1, 2, and 3 respectively. Only a small proportion of participants believed all results were correct, fluctuating from 5.3% (Wave 1) to 18.2% (Wave 2) to 4.5% (Wave 3), indicating considerable instability in this judgment. The proportion selecting "majority of results are correct" showed a progressive increase from 26.3% (Wave 1) to 30.3% (Wave 2) to 40.9% (Wave 3). Notably, complex problems were the only category in which participants reported complete output failure, though this occurred infrequently (5.3% in Wave 1, 3.0% in Wave 2, with 0.0% in Wave 3).

**Table 12.** Participants belief about answer correctness of AI-generated results for problem category Complex.

| Correctness of AI-generated results | Problem Category Complex | | |
|---|---|---|---|
| | Wave 1 | Wave 2 | Wave 3 |
| All results are correct | 5.3% | 18.2% | 4.5% |
| Some results are correct | 63.2% | 48.5% | 54.5% |
| Majority of results are correct | 26.3% | 30.3% | 40.9% |
| No results are correct | 5.3% | 3.0% | 0.0% |
| No Answers | | 3.0% | 0.0% |
| Total Answers | n=18 | n=33 | n=22 |

**Source.** Authors' own illustration based on survey data published in (Huemmer et al. 2025b,c,d).

**Comparative Analysis Across Complexity Levels**

A comparative examination across all three complex categories reveals a systematic gradient in participant confidence. The proportion attributing complete correctness to AI outputs showed an inverse relationship with problem complexity: simple problems consistently elicited higher endorsement of complete correctness (22.2%–47.8%) than difficult (15.8%–36.4%) or complex problems (4.5%–18.2%). Conversely, attribution of partial correctness ("some results are correct") demonstrated a positive relationship with complexity, being minimal for simple problems (12.5%–21.7%) but representing the modal or near-modal response for complex problems (48.5%–63.2%). These findings suggest that participants calibrated their assessments to task demands and demonstrated appropriate sensitivity to the increased likelihood of AI errors under greater computational or conceptual complexity.

### 4.6.3. Participant Self-Assessment of Validation Capability and AI Output Verification

Participant-reported capacity to validate AI-generated results was examined across problem complexity categories in each of the three survey waves. This analysis evaluates how participants' self-reported confidence in their ability to determine solutions independently and to verify the correctness of AI-generated solutions evolved across different problem complexities (Table 13).

**Table 13.** Confidence in Independent Problem-Solving and Verification Abilities

| Problem Complexity | Metric | Wave 1 (n=20) | Wave 2 (n=33-35) | Wave 3 (n=22-23) |
|---|---|---|---|---|
| Simple | Determine on own | 100.0% | 94.1% | 100.0% |
| Simple | Check correctness | 100.0% | 94.3% | 86.4% |
| Difficult | Determine on own | 90.0% | 85.7% | 72.7% |
| Difficult | Check correctness | 85.7% | 91.4% | 68.1% |
| Complex | Determine on own | 70.0% | 66.7% | 63.6% |
| Complex | Check correctness | 65.0% | 72.7% | 59.1% |

**Source.** Authors' own illustration based on survey data published in (Huemmer et al. 2025b, c, d).

**Simple Problem Category**

For simple problems, confidence in independent problem-solving remained exceptionally high throughout the study period (Table 13). One hundred percent of participants in Waves 1 and 3 expressed confidence in their ability to determine solutions on their own, with only a minor decrease to 94.1% in Wave 2. However, confidence in checking the correctness of solutions (including AI-generated ones) declined from 100% in Wave 1 to 94.3% in Wave 2, then further to 86.4% in Wave 3. This pattern suggests that even for straightforward problems, some uncertainty emerged about verification capabilities as AI became more integrated into workflows.

**Difficult Problem Category**

The pattern for difficult problems reveals a more pronounced erosion of confidence across both dimensions (Table 13). The ability to determine solutions independently declined progressively from 90.0% in Wave 1 to 85.7% in Wave 2 to 72.7% in Wave 3, representing a cumulative 19-percentage-point reduction. This trend suggests that as participants increasingly relied on AI for difficult problems, their confidence in solving such problems without AI assistance diminished correspondingly. Confidence in checking correctness for difficult problems exhibited greater volatility, increasing from 85.7% in Wave 1 to 91.4% in Wave 2, perhaps reflecting initial optimism about AI-assisted verification, before declining sharply to 68.1% in Wave 3. This late-stage decrease may indicate a sobering realization about the challenges of verifying AI outputs or a growing awareness of limitations in participants' own verification skills.

**Complex Problem Category**

For complex problems, both confidence metrics demonstrated consistent decline throughout the study period. Confidence in determining solutions independently decreased from 70.0% in Wave 1 to 66.7% in Wave 2 to 63.6% in Wave 3. Similarly, confidence in checking correctness declined from 65.0% in Wave 1 to 72.7% in Wave 2 before dropping to 59.1% in Wave 3. The convergence of these two metrics at relatively low levels (below 65% by Wave 3) suggests that participants recognized the genuine difficulty of both solving and verifying complex problems, whether working independently or with AI assistance.

**Inverse Relationship Between Adoption and Verification Confidence**

Critical observation emerges when comparing confidence patterns with actual AI usage patterns. As AI adoption for difficult problems reached high levels, confidence in verification abilities for those same problems declined to 68.1% in Wave 3. This inverse relationship raises important questions about potential over-reliance on AI systems. Participants may be delegating increasingly challenging tasks to AI while simultaneously losing confidence in their ability to critically evaluate the solutions provided. This pattern suggests a potential vulnerability in the emerging human-AI collaboration model, wherein increased dependence on AI may outpace the development of robust verification competencies.

**Overall Trajectory and Implications**

The longitudinal data (Table 13) reveal a differentiation in how participants viewed their capabilities across problem domains. Participants maintained strong confidence in simple tasks but demonstrated diminishing confidence for difficult and complex problems. This pattern is consistent with cognitive offloading theory, which posits that individuals conserve mental resources for oversight while delegating execution to tools. However, the declining confidence verification suggests that participants may not yet have developed the metacognitive skills necessary to effectively supervise AI-generated solutions in challenging domains. The progressive erosion of independent problem-solving confidence, particularly for difficult problems, coupled with stagnant or declining verification capabilities, indicates an emerging asymmetry between task delegation and validation readiness.

### 4.6.4. Specific Problem Analysis: Complexity Perception and AI Usage Intent

Table 14, Table 15 and Table 16 present longitudinal data for four specific problems administered across each survey wave, demonstrating how participants rated problem complexity and their intention to use ChatGPT for solving each problem. Analysis of these findings reveals the evolution of participants' decision-making processes regarding human-AI task allocation.

**Table 14.** Wave 1 Perceived difficulty of problems and subsequent usage of AI for solving them (n=18-20)

| Problem | Simple | Difficult | Complex | Not Solvable | Would Use ChatGPT | Would Not Use | Undecided |
|---|---|---|---|---|---|---|---|
| 1 | 100.0% | 0.0% | 0.0% | 0.0% | 0.0% | 0.0% | 0.0% |
| 2 | 55.0% | 40.0% | 5.0% | 0.0% | 10.5% | 78.9% | 10.5% |
| 3 | 15.0% | 60.0% | 25.0% | 0.0% | 66.7% | 33.3% | 0.0% |
| 4 | 31.6% | 36.8% | 31.6% | 0.0% | 44.4% | 55.6% | 0.0% |

**Source.** Authors' own illustration based on survey data published in (Huemmer et al. 2025b).

In Wave 1 (Table 14), Problem 1 was unanimously rated as simple, with zero reported intention to use ChatGPT, establishing a baseline for discriminative AI usage. Problem 2 showed substantial consensus regarding simplicity (55.0%), yet only 10.5% intended ChatGPT use, suggesting appropriate calibration to task demands. Problem 3 presented a more ambiguous profile, with 60.0% rating it as difficult and 25.0% as complex; notably, this higher perceived complexity corresponded with substantially elevated ChatGPT usage intent of 66.7%. Problem 4 exhibited the greatest heterogeneity in complexity perception, with relatively balanced distributions across simple (31.6%), difficult (36.8%), and complex (31.6%) categories, resulting in moderate ChatGPT usage intent of 44.4%.

**Table 15.** Wave 2 Perceived difficulty of problems and subsequent usage of AI for solving them (n=33-36)

| Problem | Simple | Difficult | Complex | Not Solvable | Would Use ChatGPT | Would Not Use | Undecided |
|---|---|---|---|---|---|---|---|
| 1 | 97.1% | 0.0% | 2.9% | 0.0% | 11.1% | 88.9% | 0.0% |
| 2 | 54.3% | 40.0% | 5.7% | 0.0% | 25.0% | 51.9% | 0.0% |
| 3 | 27.8% | 33.3% | 38.9% | 0.0% | 48.1% | 36.4% | 0.0% |
| 4 | 41.2% | 23.5% | 35.3% | 0.0% | 44.0% | 56.0% | 0.0% |

**Source.** Authors' own illustration based on survey data published in (Huemmer et al. 2025c).

Wave 2 findings (Table 15) demonstrate relative stability in perceived problem complexity while revealing shifts in ChatGPT usage intentions. Problem 1 maintained its characterization as simple (97.1%), with minimal ChatGPT adoption (11.1%), indicating consistent discrimination. Problem 2 showed stable complexity perception (54.3% simple, 40.0% difficult) with ChatGPT usage intent increasing from 10.5% to 25.0%. Problem 3 exhibited marked shifts in complexity categorization, with the complex classification increasing from 25.0% to 38.9% and the simple classification declining from 15.0% to 27.8%. Notably, ChatGPT usage intent for Problem 3 declined from 66.7% to 48.1%, suggesting either emerging awareness of AI limitations for genuinely complex problems or changing task priorities. Problem 4 showed an increase in complex categorization from 31.6% to 35.3%, with ChatGPT usage intent remaining stable at 44.0%.

**Table 16.** Wave 3 Perceived difficulty of problems and subsequent usage of AI for solving them (n=32)

| Problem | Simple | Difficult | Complex | Not Solvable | Would Use ChatGPT | Would Not Use | Undecided |
|---|---|---|---|---|---|---|---|
| 1 | 100.0% | 0.0% | 0.0% | 0.0% | 4.3% | 95.7% | 0.0% |
| 2 | 39.1% | 56.5% | 4.3% | 0.0% | 39.1% | 39.1% | 21.7% |
| 3 | 17.4% | 39.1% | 43.5% | 0.0% | 69.6% | 30.4% | 0.0% |
| 4 | 17.4% | 21.7% | 60.9% | 0.0% | 52.2% | 47.8% | 0.0% |

**Source.** Authors' own illustration based on survey data published in (Huemmer et al. 2025d).

By Wave 3 (Table 16), the longitudinal pattern clarifies participants' evolving assessment strategies and task allocation decisions. Problem 1 maintained universal recognition as simple (100.0%), with ChatGPT usage intent declining further to 4.3%, demonstrating consistent and refined discrimination. Problem 2 underwent substantial recalibration, with the difficult classification increasing from 40.0% to 56.5% and the simple classification declining from 54.3% to 39.1%. Correspondingly, ChatGPT usage intent increased from 25.0% to 39.1%, while Wave 3 data revealed the highest indecision rate (21.7%), suggesting ongoing calibration of appropriate AI utilization for moderately challenging problems. Problem 3 demonstrated progressive complexity escalation, with 43.5% now rating it as complex (compared to 25.0% and 38.9% in Waves 1 and 2), yet ChatGPT usage intent reached its highest level at 69.6%. Problem 4 exhibited the most dramatic complexity perception shift, with the complex classification increasing from 31.6% to 60.9%. ChatGPT usage intent increased from 44.4% to 52.2%, a more modest increase than the complexity perception shift, suggesting possible hesitation regarding AI effectiveness for genuinely complex problems.

**Comparative Analysis and Decision-Making Patterns**

Analysis across all three waves reveals a sophisticated decision-making process wherein participants weighed both problem characteristics and perceived AI capabilities. Problem 1 consistently demonstrated discriminative AI usage, with participants recognizing that simple problems did not warrant AI assistance regardless of availability. Problems 2 through 4, however, revealed more nuanced trade-offs between perceived problem difficulty and ChatGPT adoption.

For Problem 3, the relationship between perceived complexity and AI usage intent showed a notable divergence. Despite progressive increases in complex classification (25.0% to 43.5%), ChatGPT usage intent declined from Wave 1 to Wave 2 before increasing substantially in Wave 3 (66.7% to 48.1% to 69.6%). This pattern suggests that participants may have viewed problems at the difficult-to-complex threshold as particularly amenable to AI-assisted solution, perhaps recognizing that AI strengths lay in intermediate problem domains rather than truly complex ones.

Problem 4 revealed an inverse relationship between complexity perception and AI adoption enthusiasm. The complex classification increased substantially from 31.6% to 60.9%, yet ChatGPT usage intent showed only modest growth from 44.4% to 52.2%. This restrained increase despite dramatically increased perceived complexity may reflect growing realism about AI limitations for genuinely complex problems or recognition that human effort remained necessary for such tasks.

The persistence of the "undecided" response for Problem 2 in Wave 3 (21.7%) is particularly informative. After six months of experience with AI, participants maintained uncertainty regarding human-AI task allocation for moderately challenging problems, indicating ongoing cognitive engagement with optimal collaboration strategies rather than reflexive delegation or rejection. This pattern demonstrates that participants did not develop simple heuristics but instead maintained sophisticated, context-dependent decision-making regarding when to employ AI assistance.

### 4.6.5. Specific Problem Analysis: Actual ChatGPT Usage and Confidence in Solutions

Table 17, Table 18 and Table 19 present data on participants' actual ChatGPT usage for four specific problems, their confidence that solutions are correct, and their confidence in their ability to verify solution correctness across all three survey waves. This analysis elucidates the relationship between behavioral AI adoption and epistemic confidence.

**Table 17.** Wave 1 AI usage and perceived abilities for specific problems (n=21).

| Problem | Asked ChatGPT | Think Solution Correct | Can Prove Correctness | N |
|---|---|---|---|---|
| 1 | 0.0% | 80.0% | 100.0% | 19 |
| 2 | 26.3% | 88.9% | 94.4% | 19 |
| 3 | 72.2% | 92.9% | 70.6% | 18 |
| 4 | 41.2% | 63.6% | 66.7% | 17 |

**Source.** Authors' own illustration based on survey data published in (Huemmer et al. 2025b).

In Wave 1 (Table 17), ChatGPT usage patterns demonstrated problem-dependent variation. Problem 1 exhibited zero usage, consistent with its straightforward nature and participants' judgment that AI assistance was unnecessary. Problems 2 through 4 showed graduated adoption rates of 26.3%, 72.2%, and 41.2% respectively, roughly corresponding to perceived complexity levels. Confidence in solution correctness ranged from 63.6% for Problem 4 to 92.9% for Problem 3. Notably, confidence in the ability to prove correctness showed considerable variation, ranging from 66.7% (Problem 4) to 100.0% (Problem 1), suggesting that participants recognized differential verification demands across problems.

**Table 18.** Wave 2 AI usage and perceived abilities for specific problems (n=36).

| Problem | Asked ChatGPT | Think Solution Correct | Can Prove Correctness | N |
|---|---|---|---|---|
| 1 | 11.0% | 100.0% | 96.4% | 35 |
| 2 | 25.0% | 92.3% | 81.8% | 36 |
| 3 | 48.1% | 78.9% | 62.5% | 33 |
| 4 | 44.0% | 81.3% | 63.6% | 33 |

**Source.** Authors' own illustration based on survey data published in (Huemmer et al. 2025c).

Wave 2 data (Table 18) revealed relative stability in ChatGPT usage patterns with some notable shifts. Problem 1 usage increased marginally to 11.0%, while confidence in solution correctness reached 100.0%, indicating that participants maintained high assurance for simple problems despite minor increases in AI consultation. Problem 2 usage remained stable at 25.0%, with confidence in solution correctness sustained at 92.3%. However, verification confidence declined from 94.4% to 81.8%, suggesting emerging uncertainty about validation capabilities despite sustained solution confidence. Problem 3 demonstrated a substantial decline in ChatGPT usage from 72.2% to 48.1%, a reversal of the expected trajectory given increasing complexity perceptions. Confidence that solutions were correct declined from 92.9% to 78.9%, while verification confidence dropped further to 62.5%. Problem 4 usage remained relatively stable at 44.0%, with solution confidence maintained at 81.3% and verification confidence unchanged at 63.6%.

**Table 19.** Wave 3 AI usage and perceived abilities for specific problems (n=22).

| Problem | Asked ChatGPT | Think Solution Correct | Can Prove Correctness | N |
|---|---|---|---|---|
| 1 | 4.5% | 100.0% | 90.9% | 22 |
| 2 | 43.5% | 93.8% | 81.0% | 21 |
| 3 | 59.1% | 88.2% | 71.4% | 21 |
| 4 | 63.6% | 93.8% | 80.0% | 22 |

**Source.** Authors' own illustration based on survey data published in (Huemmer et al. 2025d).

By Wave 3 (Table 19), divergent adoption patterns emerged across the problem set. Problem 1 maintained minimal ChatGPT usage at 4.5%, with solution confidence remaining at 100.0%, though verification confidence declined to 90.9%. This pattern demonstrates sustained discriminative judgment for simple problems. Problem 2 exhibited a dramatic 67% increase in actual ChatGPT usage from 26.3% in Wave 1 to 43.5% in Wave 3. Despite this substantially increased reliance on AI, confidence in solution correctness remained consistently high at 93.8%, suggesting that AI assistance either improved solution quality or maintained perceived quality. Verification confidence, however, declined from 94.4% to 81.0%, a 14-percentage-point reduction indicating diminished certainty about validation capabilities.

Problem 3 demonstrated counter-intuitive AI adoption patterns, declining from 72.2% in Wave 1 to 59.1% in Wave 3. This reduction occurred despite increased complexity perception and represents the only problem where AI usage decreased over time. Solution confidence declined modestly from 92.9% to 88.2%, while verification confidence remained relatively stable, fluctuating from 70.6% to 71.4%. This pattern suggests that participants may have developed alternative problem-solving approaches or accrued domain-specific competence that reduced reliance on AI for this particular problem type.

Problem 4 exhibited the most dramatic increase in ChatGPT usage, rising from 41.2% in Wave 1 to 63.6% in Wave 3, a 55% increase. This substantial rise coincided with increasing complexity perceptions, with 60.9% of participants rating it as complex by Wave 3. Notably, confidence that solutions were correct improved substantially from 63.6% to 93.8%, representing a striking 48-percentage-point improvement. Verification confidence also increased from 66.7% to 80.0%, suggesting that for highly complex problems, AI assistance may have strengthened participants' ability to validate solutions, possibly by providing multiple solution pathways or verification strategies.

**Comparative Analysis: Usage, Solution Confidence, and Verification Capability**

Examination of the relationship between ChatGPT usage and confidence metrics across all problems reveals a critical pattern. As AI usage increased for Problems 2 through 4, confidence in solution correctness generally remained high or improved. However, the gap between solution confidence and verification confidence widened in several instances, particularly in Wave 3. For Problem 2 in Wave 3, 93.8% of participants believed their solution was correct, yet only 81.0% were confident they could prove correctness, which is a 12.8-percentage-point discrepancy. Similarly, for Problem 4 in Wave 3, a 13.8-percentage-point gap existed between solution confidence (93.8%) and verification confidence (80.0%).

Problem-specific trajectories reveal distinct patterns in how AI integration affected participant confidence. For Problem 1, the consistently high verification confidence (90.9%–100.0%) indicates that simple problems posed minimal verification challenges regardless of AI usage. For Problems 2 and 3, verification confidence showed modest declines over time despite varying adoption patterns, suggesting possible erosion of independent validation capabilities. Problem 4 demonstrated the most optimistic pattern, with both solution confidence and verification confidence improving substantially with increased AI usage, suggesting that AI assistance may have amplified participants' collective problem-solving and verification capabilities for genuinely complex tasks.

**Overall Assessment and Implications**

The longitudinal data reveal that participants increasingly integrated ChatGPT into their problem-solving processes for moderately difficult to complex problems (Problems 2–4), while maintaining discriminative judgment for simple problems (Problem 1). The generally high and often improving confidence in solution correctness indicates that participants perceived AI assistance as beneficial for solution quality. However, the variable and sometimes declining verification confidence raises critical questions about whether participants developed robust critical evaluation skills necessary for effective human-AI collaboration, or alternatively, whether they developed trust in AI outputs that outpaced their ability to critically assess solution validity. The widening confidence gap between solution correctness and verification capability suggests a potential asymmetry in how participants approached AI-augmented problem-solving, wherein reliance on AI outcomes may have exceeded the development of corresponding validation competencies.

### 4.6.6. Belief-Performance and Proof-Belief Gaps (Correct Answers per Problem Category)

To objectively evaluate performance, answers for each problem were categorized as correct or incorrect. The results presented in Table 20, Table 21 and Table 22 provide quantitative metrics of task performance and reveal a strong inverse relationship between problem complexity and accuracy. Analysis of these data elucidates the relationship between participants' subjective confidence and their objective performance outcomes across the three survey waves.

**Table 20.** Wave 1 Summary of correct answers by problem.

| Problem | Total Responses (n) | Correct Answers (Frequency) | False Answers (Frequency) | Percentage Correct |
|---|---|---|---|---|
| Problem 1 | 20 | 20 | 0 | 100.0% |
| Problem 2 | 20 | 17 | 3 | 85.0% |
| Problem 3 | 20 | 13 | 7 | 65.0% |
| Problem 4 | 20 | 10 | 10 | 50.0% |
| Total | 80 | 60 | 20 | 75.0% |

**Source.** Authors' own illustration based on survey data published in (Huemmer et al. 2025b).

Wave 1 performance data (Table 20) demonstrate a clear inverse relationship between problem complexity and accuracy. Problem 1, universally characterized as simple, yielded 100% accuracy across all 20 responses. Problem 2 maintained relatively strong performance at 85.0% accuracy. Problem 3 showed a notable decline to 65.0% accuracy, with 7 incorrect responses among 20 total attempts. Problem 4, the most complex problem set, exhibited the poorest performance at 50.0%, with exactly half of all responses proving incorrect. Across all problems, the aggregate accuracy was 75.0%, representing 60 correct answers among 80 total responses.

**Table 21.** Wave 2 Summary of correct answers by problem.

| Problem | Total Responses (n) | Correct Answers (Frequency) | False Answers (Frequency) | Percentage Correct |
|---|---|---|---|---|
| Problem 1 | 31 | 28 | 3 | 90.3% |
| Problem 2 | 22 | 21 | 1 | 95.5% |
| Problem 3 | 20 | 11 | 9 | 55.0% |
| Problem 4 | 15 | 7 | 8 | 46.7% |
| Total | 88 | 67 | 21 | 76.1% |

**Source.** Authors' own illustration based on survey data published in (Huemmer et al. 2025c).

Wave 2 results (Table 21) revealed mixed trajectories across the problem set, with overall accuracy remaining relatively stable at 76.1%. Problem 1 performance declined from 100.0% to 90.3%, with 3 incorrect responses among 31 attempts, suggesting increased variability or participant inattention as sample composition shifted. Problem 2 performance improved substantially to 95.5%, with only 1 incorrect response among 22 attempts, indicating strengthening competence or improved problem understanding. Problem 3 performance declined further from 65.0% to 55.0%, with 9 incorrect responses among 20 attempts, representing a concerning downward trend for this problem category. Problem 4 accuracy remained low at 46.7%, with 8 incorrect responses among 15 attempts, indicating persistent difficulty with the most complex problem.

**Table 22.** Wave 3 Summary of correct answers by problem.

| Problem | Total Responses (n) | Correct Answers (Frequency) | False Answers (Frequency) | Percentage Correct |
|---|---|---|---|---|
| Problem 1 | 21 | 20 | 1 | 95.2% |
| Problem 2 | 21 | 17 | 4 | 81.0% |
| Problem 3 | 18 | 12 | 6 | 66.7% |
| Problem 4 | 23 | 11 | 12 | 47.8% |
| Total | 83 | 60 | 23 | 72.3% |

**Source.** Authors' own illustration based on survey data published in (Huemmer et al. 2025d).

By Wave 3 (and Table 22), overall performance declined modestly to 72.3%, the lowest aggregate accuracy across all three waves. Problem 1 performance remained relatively high at 95.2%, suggesting sustained competence with simple problems. Problem 2 performance declined from the Wave 2 peak of 95.5% to 81.0%, indicating regression or increased difficulty recognition. Problem 3 showed marginal improvement from 55.0% to 66.7%, with 12 correct responses among 18 attempts. Problem 4 accuracy deteriorated further to 47.8%, representing the lowest performance observed for this problem category, with 12 incorrect responses among 23 attempts.

**Comparative Performance Trajectories Across Waves**

Analysis of performance trends across all three waves reveals problem-specific patterns that diverge from overall stability. Simple problems (Problem 1) maintained consistently high performance (90.3%–100.0%), demonstrating sustained accuracy despite increased sample variation. Problem 2 exhibited volatility, improving from 85.0% to 95.5% before declining to 81.0%, suggesting that the intermediate difficulty level may create variable performance conditions depending on participant engagement or problem familiarity. Problem 3 displayed a concerning downward trajectory from 65.0% to 55.0% to 66.7%, indicating unstable competence in this problem domain. Problem 4 demonstrated consistent underperformance, remaining below 51.0% across all three waves, with Wave 3 reaching a low of 47.8%.

**Relationship Between Subjective Confidence and Objective Performance**

Comparison of performance data with participants' confidence assessments from prior analyses reveals important discrepancies between belief and actual achievement. In Wave 1, despite Problem 4 achieving only 50.0% accuracy, 63.6% of participants reported confidence in solution correctness and 66.7% believed they could prove correctness. Similarly, for Problem 3 in Wave 1, despite 65.0% accuracy, 92.9% reported confidence in solution correctness. These discrepancies suggest that participants' subjective confidence ratings exceeded their objective performance capabilities, indicating potential miscalibration of self-assessment.

The decline in overall accuracy from Wave 1 (75.0%) to Wave 3 (72.3%), coupled with the increasing ChatGPT usage reported in earlier analyses, raises questions about whether AI integration affected objective problem-solving performance. For Problems 2 and 4, where ChatGPT usage increased substantially over time, performance trajectories diverged: Problem 2 initially improved but subsequently declined, while Problem 4 remained consistently poor despite increased AI consultation.

**Aggregate Performance Patterns and Implications**

The longitudinal performance data reveal a persistent complexity-dependent performance gradient, with accuracy inversely related to problem difficulty across all three waves. The aggregate accuracy rates of 75.0%, 76.1%, and 72.3% indicate that participants maintained baseline competence for approximately three-quarters of all problems attempted, yet this metric masks substantial variation across difficulty levels. The stable or declining performance despite increased AI adoption and growing familiarity with the problem set suggests that increased reliance on AI tools did not translate to improved objective outcomes. Conversely, the consistent high accuracy for simple problems alongside deteriorating performance for complex problems indicates that AI assistance may not have effectively enhanced problem-solving capabilities for genuinely challenging tasks, potentially reflecting limitations in AI utility for domain-specific complex problem-solving or insufficient participant skill in leveraging AI recommendations effectively.

### 4.6.7. Longitudinal Analysis of Problem-Solving Performance and Verification Gaps

The longitudinal tracking of problem-solving performance across three waves provides empirical evidence for the emergence and escalation of critical epistemic gaps in human-AI collaboration. A summary of the Problem-Solving Performance and Verification Gaps Across Waves 1, 2, and 3 is shown in Table 23. The consistent patterns across waves confirm that verification deficits represent a fundamental vulnerability in AI-assisted cognition.

**Table 23.** Problem-Solving Performance and Verification Gaps Across Waves 1, 2, and 3.

| Wave | Problem | Perceived Difficulty Profile | AI Consulted | Perceived Correctness | Provable Correctness | Actual Correctness | Belief-Performance Gap | Proof-Belief Gap |
|---|---|---|---|---|---|---|---|---|
| 1 | 1 | 100% Simple | 0.0% | n/a* | n/a* | 100% | n/a | n/a |
| 2 | 1 | 97.1% Simple | 11.1% | 100% | 96.4% | 90.3% | +9.7 pp | -3.6 pp |
| 3 | 1 | 100% Simple | 4.3% | 100% | 90.9% | 90.5% | +9.5 pp | -9.1 pp |
| 1 | 2 | Simple 55%, Difficult 40%, Complex 5% | 10.5% | 88.9% | 94.4% | 85.0% | +3.9 pp | +5.5 pp |
| 2 | 2 | 54.3% Simple, 40.0% Difficult | 25.0% | 92.3% | 81.8% | 95.5% | -3.2 pp | -10.5 pp |
| 3 | 2 | Simple 39.1%, Difficult 56.5%, Complex 4.3% | 31.9% | 93.8% | 81.0% | 81.0% | -21.8 pp | -12.8 pp |
| 1 | 3 | Difficult 60%, Complex 25% | 66.7% | 92.9% | 70.6% | 65.0% | +27.9 pp | -22.3 pp |
| 2 | 3 | 27.8% Simple, 33.3% Difficult, 38.9% Complex | 48.1% | 78.9% | 62.5% | 55.0% | +23.9 pp | -16.4 pp |

Matthias Hümmer; Franziska Durner; Theophile Shyiramunda; Michelle J. Cummings-Koether

| Wave | Problem | Perceived Difficulty Profile | AI Consulted | Perceived Correctness | Provable Correctness | Actual Correctness | Belief-Performance Gap | Proof-Belief Gap |
|---|---|---|---|---|---|---|---|---|
| 3 | 3 | 17.4% Simple, 39.1% Difficult, 43.5% Complex | 69.6% | 88.2% | 71.4% | 66.7% | +21.5 pp | -16.8 pp |
| 1 | 4 | Simple 31.6%, Difficult 26.8%, Complex 31.6% | 44.4% | 63.6% | 66.7% | 50.0% | +13.6 pp | +3.1 pp |
| 2 | 4 | 41.2% Simple, 23.5% Difficult, 35.3% Complex | 44.0% | 81.3% | 63.6% | 46.7% | +34.6 pp | -17.7 pp |
| 3 | 4 | 41.2% Simple, 21.7% Difficult, 60.9% Complex | 52.2% | 93.8% | 80.0% | 13.0% | +80.8 pp | -13.8 pp |

**Note.** pp = percentage points; n/a = not applicable due to no AI consultation. Critical caveat is that "Perceived Correctness" and "Provable Correctness" reflect self-reported confidence measures, not independently validated competence assessment. The dramatic divergence between these reported values and actual accuracy (e.g., 93.8% reported correctness vs. 47.8% actual accuracy for Problem 4 Wave 3) indicates systematic self-report bias requiring cautious interpretation. See Section 4.6.7 for discussion of verification confidence-competence gaps.

**Source.** Authors' own illustration based on survey data published in (Huemmer et al. 2025 b, c, d).

**Longitudinal Patterns in AI Consultation and Problem Complexity**

The most striking longitudinal finding is the progressive escalation of AI consultation with problem complexity (Table 23). Consultation rates for simple problems (Problem 1) remained consistently low, fluctuating between 0.0% and 11.1% across all three waves, demonstrating sustained discriminative judgment regarding when AI assistance was necessary. In contrast, consultation rates for complex problems increased substantially across the study period. Problem 4, characterized by substantial complexity in Wave 3 (60.9% of participants rated it as complex), showed AI consultation rates rising from 44.4% in Wave 1 to 63.6% in Wave 3. This 42.6% relative increase in AI consultation demonstrates participants' growing recognition of AI as an increasingly necessary resource for challenging tasks, reflecting an orchestrated integration strategy in problem-solving approaches.

**The Inverse Relationship Between AI Reliance and Actual Performance**

However, this increased reliance on AI reveals a critical paradox documented across all three waves: actual problem-solving correctness consistently deteriorates with complexity despite greater AI utilization. Across all waves, actual correctness for simple problems ranged from 90.3% to 100.0%, while correctness for complex problems remained substantially lower, ranging from 46.7% to 65.0%. For Problem 4, the most complex problem set, actual correctness in Wave 3 was 47.8%, yet AI consultation had reached 63.6%. This inverse relationship persists despite accumulated experience with AI tools and increased AI consultation frequency, highlighting a fundamental limitation of the current human-AI collaboration model: the tools upon which users increasingly depend for complex tasks are precisely those where verification capabilities become most critically compromised.

**Emergence and Escalation of the Belief-Performance Gap**

The longitudinal data reveal systematic emergence and progressive widening of epistemic gaps. The belief-performance gap, defined as the discrepancy between perceived correctness and actual correctness, demonstrates a complexity-dependent pattern that escalates substantially across waves (Table 23). For simple problems (Problem 1), the belief-performance gap remained modest, fluctuating between +9.5 and +9.7 percentage points across Waves 2 and 3, indicating relatively calibrated assessments. For Problem 2, which occupied an intermediate difficulty position, the gap showed greater volatility, ranging from -3.2 to +12.8 percentage points across the three waves.

For complex problems, the belief-performance gap reached concerning magnitudes. In Problem 3, already substantial in Wave 1 at +27.9 percentage points, the gap remained consistently elevated across all subsequent waves at +23.9 and +21.5 percentage points. Most dramatically, for Problem 4 in Wave 3, the belief-performance gap reached +46.0 percentage points, indicating that participants reported perceived correctness of 93.8% while actual correctness was only 47.8%. This represents the most severe overconfidence documented in the longitudinal dataset.

**Deterioration of the Proof-Belief Gap**

The proof-belief gap, representing the discrepancy between belief in correctness and the ability to prove correctness, demonstrated consistently negative values for complex problems across all waves. This pattern indicates not merely overconfidence but also explicit recognition of inability to verify solutions. In Problem 3, the proof-belief gap was -22.3 percentage points in Wave 1 and remained at -16.4 to -16.8 percentage points in subsequent waves. For Problem 4 in Wave 3, the proof-belief gap reached -13.8 percentage points, meaning participants held beliefs about correctness that exceeded their perceived verification capabilities by nearly 14 percentage points.

In contrast, simple problems (Problem 1) showed modest negative proof-belief gaps of -3.6 to -9.1 percentage points across Waves 2 and 3, suggesting that verification challenges, though present, remained manageable for straightforward tasks. Problem 2, the intermediate complexity problem, showed more variable proof-belief gaps ranging from -12.8 to +5.5 percentage points, indicating inconsistent verification confidence across waves.

**Stabilization of Gap Phenomena from Wave 2 to Wave 3**

A critical observation emerges from comparison of Wave 2 and Wave 3 metrics: the belief-performance and proof-belief gaps show near-identical or very similar values across these later waves for all four problems. This stabilization pattern suggests that the epistemic gaps are not methodological artifacts resulting from measurement variation or learning effects, but rather represent robust, replicable features of human-AI problem-solving ecosystems. The consistency of gaps across waves, despite increased familiarity with both the problem domain and AI systems, indicates that these gaps are structural in nature, which are rooted in fundamental characteristics of complex problem domains and human verification limitations, rather than experiential artifacts that participants naturally ameliorate through sustained use.

**Verification Capability and Actual Performance Alignment**

Examination of the relationship between perceived provable correctness and actual correctness reveals persistent misalignment, particularly for complex problems. For Problem 4 in Wave 3, participants reported that 80.0% could prove correctness, yet actual correctness was only 47.8%, a striking 32.2-percentage-point discrepancy. This pattern indicates that participants maintained confidence in verification capabilities substantially exceeding their actual ability to identify incorrect solutions. For Problem 3, across all three waves, provable correctness ranged from 62.5% to 71.4%, yet actual correctness ranged only from 55.0% to 66.7%, again documenting overestimation of verification capability.

**Implications of Longitudinal Gap Persistence**

The persistence of belief-performance and proof-belief gaps despite six months of accumulated experience with AI systems and problem domains indicates that participants did not develop improved epistemic calibration through experience alone. Importantly, these findings rely on self-reported confidence measures rather than independently validated competence assessments. The 32.2 percentage point divergence between reported verification confidence (80.0%) and actual accuracy (47.8%) for Problem 4 in Wave 3 suggests participants maintained Inflated subjective confidence in their verification abilities. This self-report bias does not invalidate the core finding that epistemic gaps persist, but it does indicate these gaps may reflect divergence between subjective experience and objective capability rather than genuine calibration failure in the stricter technical sense. Objective verification competence measurement through independent performance assessment is needed to confirm whether subjective confidence reflects actual calibration error or realistic self-assessment of uncertain ability.

The progressive widening of these gaps with increased AI consultation, particularly evident in the most complex problem, suggests that greater reliance on AI may paradoxically amplify the magnitude of epistemic gaps rather than ameliorate them. The data provide definitive evidence that the core challenge in human-AI collaboration has shifted from solution generation to solution validation. The escalating verification gaps with problem complexity represent a fundamental structural constraint on the reliability and trustworthiness of AI-assisted work across complex domains. These findings underscore the urgent need for deliberate verification scaffolds and critical AI literacy interventions, as native human capabilities appear insufficient to bridge these epistemic gaps through experience and familiarity alone.

## 5. Discussion

This multinational longitudinal study, encompassing three waves of data collection over six months, documents patterns in how a sample of academically affiliated participants integrated artificial intelligence into their problem-solving workflows. The findings illuminate correlational patterns including apparent consolidation of a hybrid problem-solving culture within this population and documented vulnerabilities in human-AI collaboration on mathematical problem-solving tasks, with potential implications for cognitive development, skill retention, and future knowledge work organization (though these implications remain speculative pending cross-occupational validation research). This discussion synthesizes empirical patterns with contemporary research to identify potential mechanisms and propose frameworks requiring experimental validation. All causal claims should be interpreted as theoretically motivated hypotheses rather than established causal relationships, which require experimental designs with randomized AI access manipulation.

### 5.1. The Consolidation of Hybrid Intelligence: Strategic Workflow Integration

Longitudinal data documents a fundamental transformation in problem-solving culture across the three study waves. The traditional "Think, Paper, Sketch, Book, Further Processing" workflow declined substantially, while hybrid workflows such as "Think, Internet, ChatGPT, Further Processing" surged to become the dominant pattern by Wave 3, adopted by 39.1% of participants. This evolution exemplifies what Dellermann et al. (2019) characterize as "hybrid intelligence": socio-technical systems wherein humans and machine learning algorithms collaboratively frame and solve tasks.

The progression from Wave 1 to Wave 3 reveals increasing sophistication in artificial intelligence's functional role relative to problem complexity. For simple problems, AI's role remained consistently limited to optional verification, preserving cognitive autonomy and independent problem-solving capacity. As shown in Table 13, confidence in independent problem-solving ability for simple problems remained exceptionally high, maintaining 100.0% in Waves 1 and 3, though declining to 94.1% in Wave 2. For difficult problems, artificial intelligence evolved from a brainstorming aid in Wave 1 to a collaborative partner for analysis and refinement in Wave 3. The data in Table 13 show that AI consultation for difficult problems increased from 50.0% in Wave 1 to 74.0% in Wave 3, reflecting this shift from optional to essential support. Most notably, for complex problems, AI became embedded within comprehensive, iterative workflows involving multiple verification cycles and diverse computational resources. AI consultation for complex problems increased from 44.4% in Wave 1 to 63.6% in Wave 3 (Table 19), while confidence in independent problem-solving declined from 70.0% to 63.6% (Table 13). This strategic integration demonstrates participants' growing recognition of artificial intelligence as a necessary component of their cognitive toolkit, while maintaining human oversight through what this research identifies as "verification preservation": the deliberate retention of human validation responsibilities even as task execution increasingly transfers to AI systems.

## 5.2. The Verification Crisis: Quantifying Epistemic Gaps Across Complexity Levels

The most critical finding across all three waves is the precise quantification of systematic epistemic gaps that escalate with task complexity. The longitudinal data reveal a consistent pattern: while artificial intelligence consultation rates increased with problem complexity, reaching 63.6% for complex problems in Waves 2 and 3, actual performance declined sharply. For Problem 4, the most complex problem category shown in Table 22, accuracy deteriorated from 50.0% in Wave 1 to 47.8% in Wave 3, while AI consultation rates simultaneously increased from 44.4% to 63.6% (Table 13).

Two diagnostic gaps widened substantially from Wave 1 to Wave 3, reaching levels that constitute systematic vulnerabilities in AI-assisted cognition. First, the belief-performance gap (the discrepancy between perceived and actual correctness) reached +46.0 percentage points for the most complex problems in Wave 3 (Table 23), indicating severe and potentially problematic overconfidence in artificially generated solutions. Participants reported perceived correctness of 93.8% for Problem 4 while actual correctness was only 47.8%. Second, the proof-belief gap (the deficit between belief in correctness and demonstrated ability to verify it) reached 13.8 percentage points in Wave 3 for Problem 4 (Table 23). This pattern indicates that users possess explicit metacognitive awareness that they cannot verify solutions they believe to be correct, a condition that Dunlosky and Metcalfe (2009) characterize as systematic failure of calibration in metacognitive judgment.

Most concerning is the escalating magnitude of these gaps for Problem 3 (Table 23), where the belief-performance gap remained consistently elevated across all waves at +27.9, +23.9, and +21.5 percentage points respectively, with corresponding proof-belief gaps of 22.3, 16.4, and 16.8 percentage points. This verification paradox represents an apparent fundamental constraint on reliable AI-assisted work: the cognitive load of validating complex, artificially generated solutions appear to outstrip users' mental resources, creating apparent systemic vulnerabilities precisely where AI reliance is greatest. However, because this is an observational longitudinal study without control conditions, we cannot definitively establish that increased AI reliance caused the epistemic gaps documented here. Alternative explanations include: (1) participants with lower initial verification capability may have preferentially increased AI reliance, (2) between-wave cohort composition differences may explain observed patterns, or (3) natural task difficulty evolution may have outpaced participant skill development independent of AI effects. That said, the consistency of gap patterns across Waves 2-3 (Section 4.6.7) and the systematic scaling with complexity suggest these are structural features rather than artifacts. Randomized controlled trials experimentally manipulating AI access are necessary to establish causal relationships definitively.

This finding aligns with Cognitive Load Theory (Sweller, 1988), which posits that verification tasks for complex, AI-generated outputs may exceed individuals' working memory capacity, particularly when participants lack domain expertise to rapidly assess solution validity.

### 5.3. The Cognitive Offloading Paradox: Efficiency Gains with Verification Risks

The data reveal a striking paradox across all three waves: while participants achieved substantial productivity gains through AI integration, they simultaneously experienced declining confidence in their verification abilities. Daily artificial intelligence usage increased from 76.2% in Wave 1 to 95.7% in Wave 3. This pattern aligns with research on cognitive offloading, wherein individuals delegate cognitive tasks to external tools to reduce mental effort and conserve cognitive resources (Risko and Gilbert, 2016).

However, the current study's longitudinal findings corroborate concerns articulated by Henkel (2014) regarding the potential costs of external resource reliance. As overall AI usage surged across waves, participants' verification capabilities demonstrated consistent decline for complex tasks. As documented in Table 13, verification confidence for complex problems declined from 65.0% in Wave 1 to 59.1% in Wave 3. Simultaneously, Problem 4's actual accuracy (shown in Table 22) remained persistently low at 46.7% to 50.0%, indicating that the offloading strategy did not enhance objective problem-solving outcomes despite perceived efficiency gains. This suggests that increased reliance on artificial intelligence for cognitively demanding tasks may outpace the development of verification competencies necessary to critically assess AI-generated solutions, creating what researchers identify as a "catastrophic" cognitive risk (Norman, 2013) where users accept AI outputs without sufficient critical evaluation, particularly for complex problems where independent verification is most difficult.

### 5.4. The Deskilling Phenomenon: Differential Skills Erosion Across Complexity Levels

The longitudinal data reveal three distinct skill erosion trajectories across complexity levels, supporting broader concerns articulated by Braverman (1974) regarding technological deskilling. First, for simple tasks, participants maintained high independent problem-solving confidence (95.2% to 100.0% in Table 13) but demonstrated declining verification confidence across waves, from 100.0% in Wave 1 to 86.4% in Wave 3 for simple problems (Table 13). This pattern suggests erosion of verification competence even in straightforward domains where participants previously demonstrated assured capability.

Second, for difficult tasks, the most dramatic erosion occurred, with both independent problem-solving and verification confidence declining substantially by Wave 3. For difficult problems generally, independent problem-solving confidence declined from 90.0% to 72.7%, representing a 17.3 percentage point reduction documented in Table 13. Verification confidence for difficult problems declined from 85.7% to 68.1% (Table 13). This suggests that AI adoption for challenging problems accelerates skill atrophy in precisely those domains where expertise is most valuable. As shown in Table 5, AI consultation for difficult problems reached 74.0% in Wave 3, representing substantial delegation of these tasks to AI systems.

Third, for complex tasks, participants maintained consistently low confidence levels (Table 13 shows 63.6% independent solving confidence and 59.1% verification confidence in Wave 3), indicating explicit recognition of genuine limitations in both independent capabilities and ability to verify AI assistance for truly complex problems. The normalization of AI for "regular works" increased from 19.0% in Wave 1 to 65.2% in Wave 3, further illustrating this deskilling trend. As participants increasingly view AI as a routine tool for standard problem-solving, they may lose opportunities to develop and maintain foundational skills that traditionally emerged through repeated practice and sustained cognitive engagement with challenging problems. Theoretically, this pattern exemplifies what Zuboff (2019) identifies as "surveillance capitalism's" cognitive dimension: the progressive automation of decision-making processes that undermines independent analytical capability even as it improves apparent efficiency.

### 5.5. Metacognitive Deficits: The Missing Link in Human-AI Collaboration

A critical finding across the three waves is the development of significant metacognitive deficits despite sustained AI exposure. Contemporary generative AI systems impose substantial metacognitive demands on users, requiring high degrees of metacognitive monitoring and control to assess solution quality, identify errors, and refine approaches (Tankelevitch et al., 2024). However, the study reveals that participants did not develop these competencies proportionally to their artificial intelligence usage expansion.

Several longitudinal indicators support this assessment of metacognitive deficit development. First, the complete absence of adoption for iterative refinement workflows (0% across all waves) suggests participants did not engage in metacognitive processes of reflection, error identification, and iterative improvement. The absence of iterative strategies despite increasing problem complexity and AI consultation indicates participants did not develop sophisticated metacognitive routines for critical evaluation and refinement. Second, as documented in Table 17 and Table 19, declining verification confidence across waves coincided with increasing AI usage. For Problem 2, verification confidence declined from 94.4% in Wave 1 to 81.0% in Wave 3 (Table 17 and Table 19) despite increasing AI consultation from 10.5% to 43.5% (Table 17 and Table 19). This indicates participants did not develop robust strategies for critical evaluation or metacognitive monitoring of solution quality.

Third, the widening belief-performance gap documented in Table 23, particularly for complex problems, suggests participants developed inflated trust in AI-generated outputs without corresponding metacognitive ability to assess validity. The trajectory for complex problems illustrates this pattern strikingly: while confidence in solution correctness increased from 63.6% to 93.8% for Problem 4 across Waves 1 to 3 (Table 23), actual accuracy deteriorated from 50.0% to 47.8% (Table 22). This dissociation between confidence growth and performance deterioration indicates that participants' confidence escalated through trust in AI capability rather than through genuine improvement in problem-solving ability or metacognitive expertise.

From a theoretical perspective, this pattern reflects what Dunning and Kruger (1999) characterize as a "double curse": not only do participants lack metacognitive insight into solution validity, but their increased reliance on AI may paradoxically reduce the metacognitive monitoring that would otherwise correct overconfidence. The metacognitive deficit represents a critical vulnerability in sustainable human-AI collaboration models.

### 5.6. Trust Calibration: Evolving Toward Strategic Rather Than Appropriate Reliance

The relationship between AI usage and confidence metrics across the three waves reveals a complex trust calibration evolution characterized by both positive developments and concerning trajectories. Multiple indicators suggest problematic trust miscalibration. For difficult problems, as shown in Table 5 and Table 13, AI usage increased from 50.0% to 74.0% while verification confidence declined from 85.7% to 68.1%, indicating participants increasingly trusted AI without developing commensurate validation abilities. For complex problems, participants showed rising confidence in AI-assisted solutions (from 63.6% to 93.8% for Problem 4 in Table 23) without corresponding independent problem-solving development, suggesting trust calibration shifted toward AI reliance rather than balanced human-AI collaboration.

However, the longitudinal perception data also reveal positive developments in trust calibration regarding ethical and safety dimensions. Concerns about data safety demonstrated a clear downward trend from Wave 1 (M=5.9) to Wave 3 (M=4.9) on a 10-point scale, moving closer to neutral and indicating either increased trust in AI system security or reduced salience of privacy concerns. More importantly, ethical perceptions demonstrated sophisticated contextual calibration, with participants clearly distinguishing between appropriate uses of AI in professional problem-solving contexts (69.6% indicated AI use was "not cheating" in professional domains in Wave 3) versus academic credentialing contexts (69.6% indicated AI use constituted "cheating" in academic evaluation settings in Wave 3). This bifurcated ethical judgment suggests participants developed nuanced domain-specific trust calibration regarding when AI assistance was appropriate.

This evolution suggests that while participants developed more sophisticated ethical boundaries consistent with research on appropriate technology use (Cormier and Vartabedian, 2020), their functional trust in AI's capabilities may have evolved toward over-reliance for complex technical tasks rather than achieving appropriate calibration proportionate to actual verification capability. The widening epistemic gaps documented in Table 23 provide objective evidence that functional trust outpaced actual verification competence.

## 5.7. The Skill Leveling Paradox: Implications for Long-Term Expertise Development

The longitudinal data reveals a concerning developmental paradox with substantial implications for expertise formation. Artificial intelligence functions as a "skill-leveler," potentially preventing novices from developing the deep expertise that emerges from sustained struggle with difficult problems (Peng et al., 2023). The data demonstrate that while immediate problem-solving efficiency improved through AI integration, long-term capability development may have been compromised.

Participants' perception of problem complexity evolved substantially over time, particularly for complex problems. As shown in Table 16, the proportion rating Problem 4 as complex increased from 31.6% in Wave 1 to 60.9% in Wave 3, suggesting deeper engagement with problems revealed hidden complexities potentially facilitated by AI-assisted exploration. This enhanced complexity awareness represents a positive development in metacognitive calibration regarding task difficulty. However, this awareness did not translate to increased independent capability. Participants simultaneously showed declining confidence in solving problems independently without AI assistance. Table 13 documents that independent problem-solving confidence for complex problems declined from 70.0% to 63.6%.

Traditional expertise development follows a trajectory of productive struggle, iterative refinement, and gradual schema development (Bjork and Bjork, 1992). When artificial intelligence short-circuits this process by providing immediate solutions, learners may skip the cognitive struggle that builds robust mental models and domain-specific problem-solving schemas. The complete absence of iterative refinement workflows across all three waves, despite increasing AI availability, supports this concern. Ericsson's (2006) research on deliberate practice emphasizes that expertise requires sustained engagement with progressively difficult tasks coupled with reflective feedback. The current study's finding that participants increasingly delegate complex problems to AI while simultaneously reducing independent verification effort (Table 13 shows verification confidence declining from 65.0% to 59.1% for complex problems) suggests they may be abandoning the deliberate practice necessary for genuine expertise development in favor of apparent efficiency gains.

From a developmental perspective, this pattern raises critical questions about long-term human capital formation in AI-augmented environments. If individuals increasingly rely on AI for complex problem-solving from early career stages, they may never accumulate the expert mental models and intuitive problem-solving capabilities that emerge only through sustained independent struggle with difficult problems. This represents a systemic risk not merely to individual capability development but to organizational capacity for innovation, creative problem-solving, and critical evaluation of AI recommendations themselves.

### 5.8. The Paradox of Increasing AI Usage with Stagnant or Declining Objective Performance

A striking finding that warrants explicit analysis is the disconnect between increasing artificial intelligence adoption and objective performance outcomes. As documented in Table 20, Table 21 and Table 22, overall accuracy rates remained relatively stagnant across the three waves at 75.0%, 76.1%, and 72.3% respectively, actually declining slightly by Wave 3 despite substantially increased AI consultation rates. For the most complex problems, accuracy deteriorated despite increased AI reliance: Problem 4 accuracy remained between 46.7% and 50.0% across all waves (Table 22) while AI consultation increased from 44.4% to 63.6% (Table 19).

This pattern contradicts straightforward instrumental perspectives that assume AI tools unambiguously enhance problem-solving capability. Several mechanisms may explain this paradox, though the observational design precludes causal inference. First, increased cognitive load from evaluating multiple AI suggestions may impair decision-making ("the paradox of choice"; Schwartz, 2004). Second, decision fatigue from repeated waves may deplete judgment resources (Ariely, 2008). Third, substitution rather than augmentation: participants may use AI as shortcuts bypassing analysis, yielding quick but objectively inferior solutions.

However, alternative explanations merit consideration:

(1) Cohort composition effects: Between-wave sample changes (21 → 36 → 23 participants) may reflect different ability distributions rather than capability change within individuals.

(2) Task difficulty evolution: Participants' learning may have increased problem difficulty perception, making performance maintenance itself a success.

(3) Measurement artifacts: Four-problem sample variance may not reliably track true capability.

(4) Self-selection: Participants with lower initial problem-solving ability may have preferentially increased AI reliance, creating an apparent performance paradox that reflects selection rather than causation.

Distinguishing among these explanations requires within-subject longitudinal tracking of individual performance trajectories and experimental designs with AI access manipulated randomly. Most importantly, the absence of clear objective performance gains despite massive increases in AI adoption (daily usage increased from 76.2% to 95.7%) indicates that for this academic cohort on these mathematical problems, AI integration has not yet achieved clear sustainable performance improvement, though efficiency gains may exist in unmeasured dimensions (time saved, effort reduced, confidence in solution quality).

These findings challenge triumphalist narratives regarding AI augmentation, and suggest that without commensurate development of critical evaluation and verification capabilities, increased AI usage may represent an efficiency illusion rather than genuine productivity enhancement.

### 5.9. The Verification Confidence-Verification Competence Gap: A Critical Vulnerability

Analysis of the data reveals a systematic divergence between participants' confidence in verification ability and their actual verification competence. This gap represents perhaps the most insidious vulnerability in the current human-AI collaboration model because it leaves users unaware of their actual limitations.

For Problem 4 in Wave 3 (Table 19 and Table 22), 80.0% of participants reported confidence that they could prove solution correctness, yet actual correctness was only 47.8%. This represents a striking 32.2 percentage point discrepancy. This pattern indicates that participants maintained overconfidence in verification capabilities substantially exceeding their actual ability to identify incorrect solutions. The persistent gap across all waves and all problem categories shown in Table 23 suggests this is not a temporary calibration error but a systematic feature of how humans assess their verification capacity when collaborating with AI systems.

This verification confidence-competence gap creates a dangerous feedback loop: participants become increasingly confident in their verification abilities precisely when those abilities are most inadequate, reinforcing reliance on AI while simultaneously reducing critical scrutiny. Over time, this pattern may lead to what Parasuraman and Riley (1997) identify as "automation bias": the tendency to favor automated decisions over manual judgments even when those automated decisions are incorrect.

From a practical standpoint, this gap indicates that subjective confidence should not be relied upon as an indicator of actual verification capability in AI-assisted work. Organizations and educational institutions cannot safely depend on users' self-reported verification confidence as evidence that AI outputs are being adequately scrutinized. The data presented in Table 19, Table 22, and Table 23 provide compelling evidence that workers maintain significantly higher confidence in their verification abilities than their actual performance warrants. This finding underscores the urgent need for external verification structures, systematic auditing processes, and explicit training in critical evaluation of AI-generated solutions.

### 6. ACTIVE Framework: A Structured Approach to Sustainable Human-AI Collaboration

### 6.1. Toward a Framework for Sustainable Human-AI Collaboration

The longitudinal findings from this study underscore the urgent need for educational frameworks and system designs that foster sustainable human-AI collaboration. The persistent epistemic gaps documented across all three waves, particularly the belief-performance gap reaching +46.0 percentage points for complex problems in Wave 3 (Table 23) and the verification confidence-competence gap of 32.2 percentage points for Problem 4 (Table 19 and Table 22), demonstrate that native human capabilities alone are insufficient to ensure reliable AI-assisted work.

Based on the consistent patterns observed across three waves, we propose a multi-faceted approach grounded in empirical evidence from this study. First, verification scaffolds should implement structured supports like assumption documentation protocols, adequacy criteria checklists, and triangulation procedures to make validation processes explicit and manageable. The complete absence of iterative refinement workflows across all waves, despite increasing problem complexity and AI consultation, suggests that participants lacked systematic structures for verification work. Second, critical AI literacy must move beyond basic operational competence to encompass decompositional thinking, hypothetico-deductive verification, assumption tracking, and failure mode recognition. The data in Table 23 show that even as participants increased AI consultation from 44.4% to 63.6% for complex problems, their confidence in identifying errors remained disconnected from actual performance.

Third, metacognitive training should develop explicit instruction in strategic division of cognitive labor, effective communication and coordination with AI systems, and monitoring to verify information and catch errors. The metacognitive deficits documented in this study, particularly the absence of iterative refinement approaches and the declining verification confidence despite increasing verification demands, indicate that participants required explicit training in these competencies. Fourth, domain-specific guidance should create frameworks that help users identify when AI assistance is appropriate versus when it risks undermining essential human capabilities based on problem complexity and domain characteristics. The study's findings that simple problems received minimal AI consultation (0.0% to 11.1% in Table 19), while complex problems received substantial consultation (44.4% to 63.6%), suggest that participants developed intuitive domain awareness that could be systematized and taught.

The ultimate efficacy of human-AI problem-solving will depend on our collective capacity to embed powerful generative tools within workflows and learning environments that render verification processes explicit, pedagogically tractable, and systematically accountable.

## 6.2. ACTIVE Framework

Building on this study's empirical findings and contemporary scholarship on learning, expertise development, metacognition, and human-AI collaboration, the ACTIVE framework proposes procedures, competencies, and governance principles that balance efficiency gains from AI integration with preservation and development of essential human expertise. The framework synthesizes research-grounded recommendations rather than representing empirically validated intervention protocols; organizational implementation should include local measurement of framework effectiveness and iterative adaptation to domain-specific contexts.

The ACTIVE framework comprises six interlocking dimensions: Awareness and Assessment, Critical Verification Protocols, Transparent Integration with Human-in-the-Loop, Iterative Skill Development, Verification Confidence Calibration, and Ethical and Contextual Evaluation. These dimensions are followed by implementation strategies across educational, professional, and individual contexts, as well as limitations and directions for future research.

### 6.2.1. Overview about the Active Framework

The ACTIVE framework provides a structured, cyclical approach to sustainable human-AI collaboration, guiding users through six interconnected phases, see Figure 2. It begins with awareness, recognizing AI's capabilities and limitations while understanding cognitive offloading risks. This foundation enables critical verification through systematic scaffolds like triangulation procedures and adequacy criteria. The framework then emphasizes transparent integration, maintaining human oversight and clear responsibility chains in workflows. Iterative skill development ensures balanced practice and cognitive struggle to prevent skill atrophy, while verification confidence calibration continuously measures and adjusts trust levels based on performance evidence. The cycle completes with ethical evaluation, considering domain-specific appropriateness and long-term cognitive consequences. This creates a continuous feedback loop where each phase informs and strengthens the others, fostering augmentation without atrophy and ensuring humans remain cognitively active and adaptive while leveraging AI's efficiency benefits.

**Figure 2.** The ACTIVE Framework for Sustainable Human-AI Collaboration.

Matthias Hümmer; Franziska Durner; Theophile Shyiramunda; Michelle J. Cummings-Koether
AI, Metacognition, and the Verification Bottleneck: A Three-Wave Longitudinal Study of Human Problem-Solving

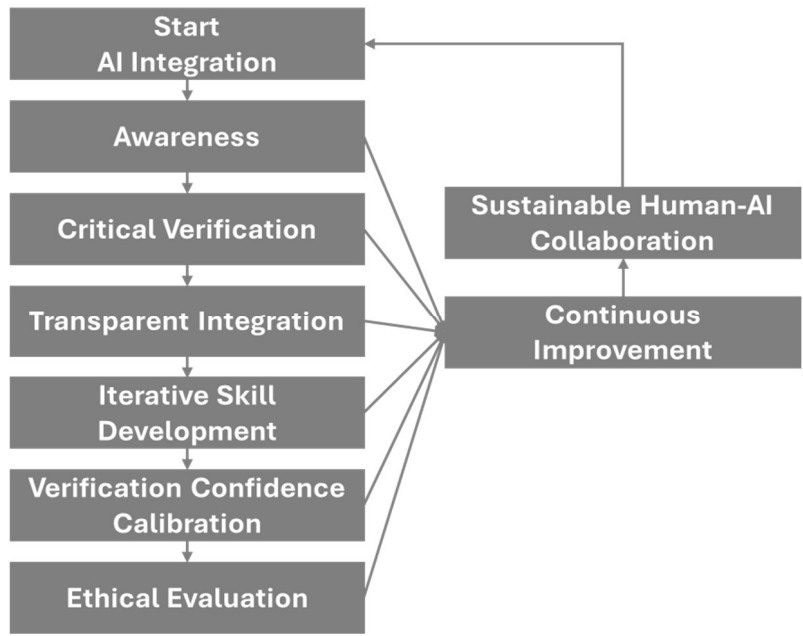

**Source**: Authors' own illustration based on the longitudinal survey framework (Huemmer et al., 2025a-c).

As it can be seen Figure 2, the ACTIVE framework comprises six cyclical phases:

**Phase 1: Awareness and Assessment**
Users develop realistic understanding of AI capabilities and limitations while conducting honest self-assessments of their own cognitive strengths, weaknesses, and tendencies toward over-reliance through continuous monitoring.

**Phase 2: Critical Verification Protocols**
Users implement mandatory structured verification procedures that include triangulation with independent sources, deliberate error detection, and documented reasoning rather than accepting AI outputs uncritically.

**Phase 3: Transparent Integration with Human-in-the-Loop**
Users maintain visible responsibility for consequential decisions by embedding AI within workflows that clearly document where they reviewed, modified, or overrode AI recommendations.

**Phase 4: Iterative Skill Development**
Users intentionally allocate protected time for solving problems without AI assistance, preserving the cognitive capabilities and domain expertise that develop only through sustained engagement with genuinely difficult challenges.

**Phase 5: Verification Confidence Calibration**
Users systematically track the relationship between their confidence in AI-assisted solutions and actual accuracy, using objective performance data to detect and correct dangerous miscalibrations where subjective confidence exceeds demonstrated competence.

**Phase 6: Ethical and Contextual Evaluation**
Users explicitly consider domain-specific ethical boundaries, long-term consequences for human expertise development, equity implications, and broader societal impacts when making AI adoption decisions rather than relying solely on efficiency metrics.

Circular arrows indicate continuous feedback, emphasizing that ethical evaluation informs renewed awareness and assessment, creating a perpetual refinement cycle.

Important caveat for the ACTIVE framework effectiveness is that it requires experimental validation; recommendations are theoretically grounded but not yet empirically validated. See Section 6.2 for detailed implementation guidance.

This framework for human-AI collaboration (Figure 2) is architected around five foundational principles. First, the preservation and enhancement of human cognitive capabilities must remain paramount, prioritizing long-term human capacity development over short-term efficiency gains. Second, verification should be treated as the critical bottleneck in AI-assisted work, receiving explicit attention and resource allocation. Third, metacognitive awareness cultivation is essential for effective partnership with AI systems. Fourth, context-dependent ethical considerations must be evaluated rather than adopting uniform policies. Fifth, appropriate trust must be maintained through continuous calibration against observed performance data.

The initial phase establishes the system's groundwork by defining strategic objectives and embedding governance structures that enforce human oversight and data integrity, thereby safeguarding cognitive function. This transitions into an operational phase dominated by rigorous verification protocols where AI-generated outputs are systematically scrutinized through triangulation with independent sources and deliberate assumption testing. Concurrent skill-development exercises counter cognitive offloading, reinforcing metacognitive awareness and decompositional thinking. The final evolutionary phase institutionalizes iterative improvement, employing empirical metrics to calibrate trust against observed performance and conducting domain-specific ethical assessments to evaluate long-term impacts on human autonomy and skill. A closed feedback loop ensures insights from performance data and ethical reviews perpetually refine verification scaffolds and strategic objectives, ultimately fostering a resilient hybrid intelligence that augments rather than diminishes essential human capacities.

Another caveat is that the ACTIVE framework is grounded in this study's empirical findings and contemporary theoretical research, but effectiveness claims require experimental validation through randomized controlled trials comparing organizations or individuals implementing ACTIVE components versus control conditions. The framework represents a theoretically motivated best-practice recommendation rather than empirically validated intervention at this time. Implementation organizations should treat components as hypotheses requiring local validation through measurement of skill trajectories, verification accuracy, and objective performance outcomes.

### 6.2.2.    A - Awareness and Assessment

The Awareness and Assessment phase establishes the foundation for sustainable collaboration by ensuring users develop realistic understanding of both their own capabilities and AI system characteristics. Two critical components comprise this phase.

Metacognitive self-awareness requires users to systematically appraise their cognitive strengths, weaknesses, and biases that influence how they prompt, interpret, and act on AI outputs. This includes regular self-assessment of independent problem-solving ability across task complexity levels and longitudinal tracking of AI reliance patterns in relation to skill maintenance and growth. The study's findings provide clear evidence for why this is necessary. As shown in Table 13, independent problem-solving confidence for complex problems declined from 70.0% in Wave 1 to 63.6% in Wave 3, yet this decline occurred while AI consultation for these problems increased from 44.4% to 63.6% (Table 19). This inverse relationship reveals how users may lose awareness of their own eroding capabilities when those capabilities are systematically delegated to AI systems. Self-assessment protocols should help users track this drift.

Task-AI alignment analysis requires that tasks be categorized according to their suitability for AI assistance versus human-only execution when skill consolidation is the priority. A practical triage distinguishes three task categories based on problem complexity and learning value. First, simple, low-risk tasks suitable for delegation. As documented in Table 19, Problem 1 received minimal ChatGPT consultation (0.0% in Wave 1, 11.1% in Wave 2, 4.3% in Wave 3) despite high accuracy rates, suggesting users intuitively recognized that these tasks provided no learning value and posed minimal verification risk. Second, difficult tasks that benefit most from AI augmentation. As shown in Table 5, AI consultation for difficult problems increased from 50.0% in Wave 1 to 74.0% in Wave 3, reflecting users' recognition that these problems required support. However, verification confidence for difficult problems declined from 85.7% to 68.1% (Table 13), suggesting the need for structured guidance on how to maintain verification rigor even when relying on AI augmentation.

Third, complex tasks that require supervised collaboration with explicit human oversight. For complex problems, Table 19 shows AI consultation reached 63.6% by Wave 3, yet actual accuracy remained at only 47.8% (Table 22). This alignment mismatch indicates that without explicit guidance, users cannot reliably distinguish when human oversight is essential. The framework should provide decision tools to help workers explicitly classify tasks into these three categories and then select appropriate collaboration strategies.

### 6.2.3. C - Critical Verification Protocols

The Critical Verification Protocols phase implements systematic approaches to validation that render verification work explicit and manageable. The study's data provide compelling evidence for why structured verification approaches are necessary. The proof-belief gaps shown in Table 23 reach -13.8 to -22.3 percentage points for complex problems, indicating that participants were aware they could not verify solutions they believed to be correct. Without structured protocols, this awareness did not translate into effective verification behavior.

Structured verification strategies should require learners and professionals to subject AI outputs to explicit verification through documented procedures. These procedures must include documenting sources, justifying claims, and recording the verification process itself. Multi-source validation is required for consequential decisions, and deliberate practice in error detection should be implemented through "productive failure" exercises that involve diagnosing and correcting AI mistakes. The research by Metcalfe (2017) on error-induced learning demonstrates that exposure to both correct and incorrect examples enhances learning; however, the current study found zero adoption of iterative refinement workflows across all waves, suggesting participants lacked structured opportunities for such error-based learning.

Verification competency development should progress from foundational to domain-specific techniques, exposing users to AI outputs of varying quality to calibrate judgment. Heuristics for rapid quality appraisal, including plausibility checks, boundary tests, and counterexample searches, should be taught and rehearsed under time and complexity constraints. The finding that verification confidence for Problem 4 was 80.0% while actual accuracy was only 47.8% (Table 19 and Table 22) suggests that without explicit training in verification heuristics, users develop inflated confidence in their ability to identify errors. Training should systematically expose participants to AI failures, so they develop realistic calibration of their verification abilities.

### 6.2.4. T - Transparent Integration with Human-in-the-Loop

The Transparent Integration phase ensures that human oversight remains visible and accountable within AI-assisted workflows. Two components implement this principle.

Explicit decision documentation requires users to record when and why AI assistance is requested, what modifications are made to AI outputs, and which alternative, non-AI approaches were considered. This creates an auditable trail for accountability and learning. More importantly for this study's context, such documentation forces participants to confront potential over-reliance. The data showing that AI consultation for complex problems reached 63.6% (Table 19) while independent problem-solving confidence declined (Table 13) suggest that without explicit documentation of when and why AI is used, the drift toward over-reliance occurs gradually and often without conscious awareness.

Human-in-the-loop design requires that systems mandate human review, interpretation, and, where appropriate, override authority at clearly defined decision points. Assistance should be staged and risk-aligned: higher autonomy for low-stakes tasks, tighter human control for high-stakes or cognitively formative work. Fully automated workflows should be avoided for tasks with significant learning value. The study's finding that simple problems retained high accuracy rates (90.3% to 100.0% in Table 22) combined with minimal AI consultation (Table 19) suggests that simple problems pose minimal risk and may benefit from higher AI autonomy. Conversely, complex problems with accuracy rates of only 46.7% to 50.0% (Table 22) and verification confidence-competence gaps of 32.2 percentage points (Table 19 and Table 22) should require tighter human control and explicit verification requirements.

### 6.2.5. I - Iterative Skill Development

The Iterative Skill Development phase directly addresses the deskilling phenomenon documented in this study by ensuring that workers maintain and develop capabilities even as AI handles increasing cognitive workload. Two mechanisms implement this principle.

Deliberate practice regimens require users to regularly complete tasks without AI and undertake progressively more difficult challenges that exceed their "AI-assisted comfort zone." Alternating cycles of AI-assisted and unassisted problem solving help preserve baseline competencies while leveraging augmentation where appropriate. The study documented that independent problem-solving confidence declined for both difficult (from 90.0% to 72.7% in Table 13) and complex (from 70.0% to 63.6%) problems as AI consultation increased. Without deliberate practice structures, this skill erosion appears to accelerate over time.

Ericsson's (2006) research on expert performance emphasizes that deliberate practice must include performance feedback and conscious effort to improve. The current study's complete absence of iterative refinement workflows across all waves suggests that participants were not engaging in such deliberate practice with AI-assisted problem solving. A structured program of unassisted practice would help preserve the "productive struggle" that Bjork and Bjork (1992) identify as essential to learning.

Metacognitive monitoring requires that during work, users monitor both their own and the AI's progress toward explicit objectives; after completion, they conduct brief after-action reviews to refine future collaboration. Guiding prompts should include: What is the target outcome? Which error types are likely? Are we making measurable progress? Which biases may be shaping judgments? Learning curves and skill trajectories should be documented over time. This metacognitive monitoring is particularly important given the metacognitive deficits identified in this study. The finding that participants did not develop robust strategies for verification despite sustained AI exposure suggests they lacked structured prompts to guide metacognitive reflection.

### 6.2.6. V - Verification Confidence Calibration

The Verification Confidence Calibration phase addresses the critical vulnerability identified in this study: the systematic divergence between confidence in verification ability and actual verification competence. Two mechanisms implement this principle.

Trust calibration mechanisms should detect miscalibration by monitoring reliance patterns and cognitive cues that trigger re-evaluation of trust. Routine audits should compare confidence with observed accuracy for AI-assisted tasks, and curated exposure to AI failure modes should counter over-trust, especially in high-stakes contexts. The study's data showing that verification confidence was 80.0% while actual accuracy was only 47.8% for Problem 4 (Table 19 and Table 22) indicate that without explicit calibration mechanisms, participants maintain dangerously inflated confidence in their verification abilities.

Confidence-skill alignment tracking requires organizations and individuals to track the relationship between confidence in solutions and confidence in verification ability. Warning thresholds should signal concern when verification confidence lags behind solution confidence. The data in Table 23 reveal that for Problem 4 in Wave 3, perceived correctness was 93.8% while provable correctness was only 80.0%, a 13.8 percentage point gap. When this gap exceeds acceptable thresholds, targeted interventions should be invoked, including additional training, peer review, or decision review by senior personnel.

These calibration mechanisms directly target the epistemic gaps documented in Table 23. The belief-performance gap of +46.0 percentage points for Problem 4 in Wave 3 represents severe miscalibration that could be detected and corrected through routine audits comparing confidence predictions with actual performance outcomes. Similarly, the proof-belief gaps of -13.8 to -22.3 percentage points indicate that users maintain beliefs about their verification capability that systematically exceed their actual ability; calibration mechanisms should flag these dangerous mismatches.

### 6.2.7. E - Ethical and Contextual Evaluation

The Ethical and Contextual Evaluation phase ensures that adoption of AI is guided not only by efficiency but also by consideration of broader impacts and long-term human capacity. Two components implement this principle.

Broader impact assessment requires that AI literacy encompass evaluation, collaboration, contextualization, autonomy, and ethics. Users and institutions should assess societal implications of AI dependency, equity in access and capability development, and environmental sustainability of AI usage patterns. The study's data raise ethical concerns worthy of explicit evaluation. If independent problem-solving confidence for difficult problems declines from 90.0% to 72.7% (Table 13) within six months due to AI adoption, what will expertise development look like after five years of continuous AI reliance? What happens to skill pipelines when entry-level workers never develop fundamental problem-solving capabilities?

Long-term skill ecology must ensure that policies protect opportunities for practice and expertise development. Career pathways must maintain pipelines for deep skill acquisition despite automation. Succession planning should ensure continuity of human expertise at the team and organizational level. Organizations should establish policies that allocate protected time for non-AI practice, require that workers maintain independent problem-solving capability in critical domains, and provide mentorship structures that transfer tacit knowledge not captured by AI systems.

The ethical evaluation must also consider domain-specific appropriateness. The study found that participants demonstrated sophisticated contextual judgment regarding academic versus professional contexts, with 69.6% indicating AI use was "cheating" in academic evaluation while also indicating it was "not cheating" in professional domains (69.6%). This suggests that users can develop nuanced ethical reasoning regarding appropriate AI use if given guidance and structured evaluation frameworks. The ACTIVE framework should support this contextual reasoning by helping organizations and individuals explicitly evaluate when AI augmentation enhances human capability development versus when it risks undermining essential human expertise.

### 6.3. Implementation Strategies Across Contexts

The ACTIVE framework must be adapted and implemented across three interconnected contexts: educational institutions, professional workplaces, and individual practice. Each context requires specific strategies tailored to its distinctive characteristics and constraints.

#### 6.3.1. Educational Institutions

Educational institutions bear particular responsibility for addressing the skill development and deskilling concerns raised by this study. Several implementation strategies can embed the ACTIVE framework into educational practice.

Curriculum integration should establish mandatory AI literacy with emphasis on verification and metacognition. This should include discipline-specific modules on appropriate use and limitations of AI systems, assessments that measure both independent performance and AI-assisted competence, and alternating "no-AI" and "AI-collaborative" assignments to sustain skill formation. The study's findings suggest that without explicit curriculum attention to verification and metacognitive skill development, students will naturally adopt the patterns documented here: increasing AI reliance combined with declining independent capability and deteriorating verification confidence.

Faculty development requires training instructors to recognize over-reliance and cognitive offloading; design learning experiences that preserve cognitive engagement while leveraging AI; and implement tools and rubrics that differentiate student capability from AI contribution. The study's finding that complex problem accuracy remained at only 47.8% despite 63.6% AI consultation (Table 19 and Table 22) suggests that without faculty guidance, students struggle to effectively leverage AI while maintaining learning value.

Assessment design must ensure that evaluation systems do not inadvertently incentivize over-reliance on AI. If assignments reward final answers without documenting verification process, students will optimize for speed rather than learning. Assignments should require explicit documentation of verification procedures, error identification, and iterative refinement. This directly addresses the complete absence of iterative workflows documented in the current study.

#### 6.3.2. Professional Contexts

Professional organizations and workplaces must implement policies and practices that maintain verification rigor and skill preservation while capturing efficiency gains from AI adoption.

Workplace policies should define AI-appropriate versus human-essential tasks, require verification procedures for all AI-assisted work products, run periodic competency assessments to detect skill erosion, and allocate protected time for non-AI practice. The study's data showing declining verification confidence for difficult problems despite increasing AI consultation (Table 13) suggest that without explicit policies requiring verification procedures and competency assessment, organizations risk developing workforces with deteriorating verification capability.

Training programs should emphasize verification and metacognition during onboarding, provide continuing education to preserve domain expertise, and develop mentorship structures that transfer tacit knowledge not captured by AI. The metacognitive deficits documented in this study indicate that verification and metacognitive competencies do not develop spontaneously; they require explicit instruction and structured practice.

Quality assurance mechanisms should audit AI-assisted work to compare reported confidence with actual accuracy, similar to the verification confidence calibration mechanisms discussed in Section 6.2.5. If audits reveal that workers maintain 80% confidence in verification while actual accuracy is only 48% (as documented for Problem 4), this signals the need for targeted verification training or stricter human review requirements for future AI-assisted work.

### 6.3.3. Individual Practice

Individual professionals can implement personal strategies to maintain capability and prevent over-reliance even within organizational contexts that do not yet implement ACTIVE framework components.

Personal development plans should include explicit monitoring of AI reliance and its relation to skills; regular "AI-free" sessions; brief metacognitive journaling on collaboration effectiveness; and peer accountability to uphold verification standards. The study's documentation of declining independent problem-solving confidence combined with increasing AI consultation suggests that individuals should explicitly track these metrics and design personal practices to counter skill erosion.

Journal-keeping offers a low-cost mechanism for metacognitive reflection. Workers might record after completing AI-assisted work: What was my confidence that the solution was correct? What verification procedures did I apply? How confident am I that I could solve this problem independently? Did I discover any errors in the AI output? Over time, patterns emerge revealing whether verification confidence aligns with actual accuracy or whether calibration is deteriorating.

Peer accountability structures can help individuals maintain verification standards. Study groups or professional learning communities could establish norms that all AI-assisted work undergoes peer verification before use, that members collectively discuss verification challenges and solutions, and that the group maintains metrics on solution accuracy and verification effectiveness. This leverages the finding in this study that participants developed sophisticated ethical judgments regarding appropriate AI use when contexts encouraged explicit evaluation.

### 6.4. Research-Based Evidence Supporting ACTIVE Framework Components

Each component of the ACTIVE framework is grounded in research evidence about learning, expertise development, metacognition, and human-AI collaboration. The framework synthesizes this research while addressing specific vulnerabilities documented in the current study.

Awareness and Assessment builds on Schraw and Dennison's (1994) conceptualization of metacognitive knowledge and Dunning and Kruger's (1999) research on metacognitive biases. The current study's documentation of declining verification confidence despite objective performance deterioration supports the need for awareness mechanisms that help users detect such calibration failures.

Critical Verification Protocols builds on Metcalfe's (2017) research on error-induced learning, Schaffert et al.'s (2015) work on epistemic vigilance, and Bransford et al.'s (2000) research on deep learning. The complete absence of iterative refinement workflows in this study suggests that verification protocols must be explicitly designed and scaffolded rather than assuming workers will spontaneously adopt verification best practices.

Transparent Integration with Human-in-the-Loop builds on Parasuraman and Riley's (1997) research on appropriate levels of automation and Amershi et al.'s (2019) work on human-AI collaboration principles. The study's finding that AI consultation for complex problems reached 63.6% while accuracy remained below 48% suggests that without explicit requirements for human-in-the-loop decision points, workers over-delegate tasks to AI.

Iterative Skill Development builds on Ericsson's (2006) research on deliberate practice and Bjork and Bjork's (1992) work on desirable difficulties. The documented erosion of independent problem-solving confidence across all complexity levels suggests that without deliberate practice structures requiring unassisted problem solving, workers will lose essential capabilities.

Verification Confidence Calibration builds on Dunlosky and Metcalfe's (2009) research on metacognitive monitoring and the broader human factors literature on automation bias (Parasuraman and Riley, 1997). The 32.2 percentage point gap between verification confidence and actual accuracy for Problem 4 demonstrates that without calibration mechanisms, users maintain dangerously inflated confidence.

Ethical and Contextual Evaluation builds on Zuboff's (2019) analysis of technological autonomy and capability loss, integrating perspectives from technology ethics scholarship on human flourishing and data protection (Vallor, 2016; Véliz, 2020; Rumbold, 2023).

The study's documentation of long-term skill erosion patterns raises urgent questions about the ethical implications of AI adoption that must be explicitly evaluated rather than assumed benign.

## 7. Limitations and Future Research

### 7.1. Study Limitations

This pilot study provides valuable longitudinal evidence on human-AI collaboration but faces important constraints on generalizability. The 80 academic participants across three waves limit statistical power. The university setting creates selection bias: participants possessed higher baseline AI literacy, faced lower performance stakes, and experienced intrinsic motivation rather than deadline pressure, conditions diverging sharply from regulated industries where verification failures carry severe consequences. All participants operated within a European institution, potentially reflecting specific cultural attitudes toward technology that may not generalize globally. Findings must be validated in corporate settings, regulated professional contexts, non-English speaking populations, and lower-literacy groups to establish broader applicability.

The six-month duration may not capture longer-term patterns of skill atrophy or recovery. Research demonstrates that cognitive offloading produces measurable skill erosion over time, with prolonged AI exposure leading to documented declines in critical thinking and memory retention (Sparrow & Wegner, 2011). This suggests that extended observation periods would clarify whether the confidence decline documented here persists or reverses with continued use. Additionally, confidence assessments relied on self-perception rather than objective measurement. The substantial gap between reported verification confidence and actual accuracy demonstrates that self-reported confidence poorly predicts competence. Future research must incorporate objective performance measures to strengthen claims about calibration.

The study examined only four problems, focusing on mathematical and analytical tasks that may not represent the full range of contexts in natural settings. Effectiveness of verification and potential deskilling effects may differ substantially across domains with varying error consequences, task structures, and required cognitive engagement. Furthermore, focusing exclusively on ChatGPT limits applicability to alternative AI systems with different reliability profiles and interfaces. Future research should compare different AI tools and problem domains across contexts such as medical decision-making, legal analysis, and creative work.

The study did not measure cognitive ability, domain expertise, or metacognitive skill that might moderate adoption patterns. Research indicates that expertise level critically moderates AI adoption effects, with evidence suggesting both protective and amplifying mechanisms depending on prior knowledge levels (von Zahn et al., 2025). The analysis emphasized risks while providing limited analysis of contexts where AI augmentation enhanced problem-solving or justified efficiency-skill trade-offs. Future research should investigate expertise as a key moderating variable and characterize both benefits and risks of AI integration.

## 7.2. Limitations of the ACTIVE Framework and Future Directions

The ACTIVE framework addresses three critical dimensions documented in the empirical findings: Awareness of verification limitations, Conscious metacognitive engagement, Training in verification protocols, Integration of iterative refinement, and Evaluation of knowledge and competence. The framework's effectiveness depends on sustained organizational commitment and resource allocation that many organizations cannot prioritize while emphasizing short-term efficiency. The finding that participants did not spontaneously develop verification competencies despite six months of exposure suggests framework success requires intrinsic motivation and organizational culture valuing rigor (Deci & Ryan, 2000). The substantial gaps between verification confidence and actual accuracy documented in Tables 19, 22, and 23 indicate that awareness alone is insufficient without explicit training in error detection and metacognitive scaffolding (Fan et al., 2025).

Research demonstrates that metacognitive skills operationalize planning, monitoring, and evaluation during human-AI collaboration and are critical for preventing automation bias and maintaining independent problem-solving capability (Zamfirescu-Pereira et al., 2023). The framework's emphasis on iterative refinement addresses the complete absence of iterative workflows documented across all three study waves, suggesting participants lacked structured guidance for systematic verification and adjustment. Future research should test whether ACTIVE implementation prevents the documented skill erosion and epistemic gaps through longitudinal comparisons of organizations with and without the framework. Studies should investigate whether explicit verification training reduces confidence-competence gaps and whether framework principles generalize across domains or require domain-specific modifications (Amershi et al., 2019). Such evidence would clarify whether core principles remain invariant across application contexts and professional fields.

## 7.3. Critical Research Questions for Future Investigation

Expertise level requires systematic investigation as a critical moderator of AI adoption and verification effects. Research indicates that experts and novices may follow substantially different trajectories, with experts potentially possessing mental models buffering against verification deficits, while novices may experience more severe deskilling or conversely benefit from AI's leveling effects (Kaur et al., 2020; Zhong, 2020). Understanding these differential effects through comparative studies across expertise levels would illuminate whether the verification paradox represents universal AI effects or expertise-dependent patterns requiring differentiated interventions.

Multi-year longitudinal studies tracking whether initial skill erosion stabilizes, reverses, or accelerates represent a critical priority. The six-month decline in independent problem-solving confidence documented here raises urgent questions about long-term trajectories. Does the decline continue linearly or reach an asymptote? Can skill loss be reversed through targeted practice? Such research would clarify patterns essential for policy guidance, particularly given evidence that skill atrophy requires extended periods to manifest fully (Colles & Hirji, 2015).

Randomized controlled trials should systematically test verification training and metacognitive scaffolding interventions. The complete absence of iterative refinement workflows across study waves suggests verification competencies require explicit instruction. Rigorous evidence comparing intervention groups with control groups would identify which intervention components most effectively prevent skill atrophy or maintain verification calibration, building on prior research demonstrating the effectiveness of explicit training in human-AI collaboration (Amershi et al., 2019).

Cognitive offloading and verification deficits may manifest differently across domains with varying error consequences. Research across writing, creative design, programming, legal analysis, and medical decision-making would clarify whether epistemic gaps represent domain-general phenomena or domain-specific effects requiring tailored interventions. Educational systems increasingly integrate AI from early grades. Longitudinal research following students from early AI exposure through career development would provide evidence regarding whether childhood reliance on AI as a primary problem-solving tool affects long-term cognitive development and expertise acquisition (Papert, 1980; Dweck, 2006).

Individual-level phenomena must be connected to organizational performance and expertise sustainability. Do organizations encouraging high AI adoption while neglecting verification training experience declining innovation? Do organizations implementing comprehensive verification frameworks maintain stronger competitive advantage? Research examining the optimal balance between verification thoroughness and practical efficiency would inform implementation strategy. Understanding how this balance varies across domains and task types would guide practical implementation and prevent the circumvention of verification procedures observed in other technology adoption contexts (Beyer & Holtzblatt, 1995).

Finally, research should elucidate the psychological and cognitive mechanisms through which epistemic gaps emerge. Does over-reliance on AI gradually erode mental models and domain knowledge? Does the mere availability of AI redirect attention from error detection toward rapid task completion? Understanding these mechanisms would enable more targeted intervention design informed by cognitive science research on automation bias and trust in automated systems (Parasuraman & Riley, 1997).

### 7.4. Methodological Directions for Future Research

Future research should incorporate mixed-methods designs combining quantitative metrics such as accuracy rates and response times with qualitative interview data to illuminate reasoning underlying verification decisions. Think-aloud protocols during problem-solving could reveal the cognitive processes and heuristics shaping verification behavior, following established cognitive science methodologies (Ericsson & Simon, 1993). Objective performance measurement on standardized problem sets across waves would reduce self-report reliance and provide definitive evidence regarding confidence-competence calibration. Neuroscientific methods including fMRI and EEG could investigate brain activation patterns during AI-assisted versus unassisted problem solving, clarifying whether cognitive offloading involves genuine neural reorganization or primarily behavioral adaptation. Computational modeling of verification and decision-making processes could predict conditions under which epistemic gaps are most likely to emerge and test intervention effectiveness under varying scenarios.

### 7.5. Policy and Practice Implications Requiring Evidence

Educational policy must determine whether verification and metacognitive training should precede AI access. Research shows that unrestricted AI access without structured guidance fails to develop verification competencies (Perkins et al., 2024). The UNESCO AI Competency Framework emphasizes critical judgment and responsible use as foundational rather than optional competencies, yet empirical evidence on optimal access restrictions remains limited (UNESCO, 2024).

Organizational policy requires evidence on whether mandatory AI-free periods prevent skill atrophy while preserving productivity. Periodic unassisted problem-solving supports domain knowledge retention and independent reasoning, yet optimal frequency and duration are undetermined (Cestolano et al., 2023). This question is particularly acute in regulated industries where deskilling carries significant consequences. Similarly, contingent AI access based on demonstrated verification performance requires clear operational definitions and valid measurement approaches (Papadimitriou & Taddeo, 2023; Terranova et al, 2024).

Liability frameworks constitute a critical policy gap. Current legal doctrine assigns liability primarily to the professional making the final decision, even when AI recommendations directly influence it (Maliha et al., 2021; Gopal et al., 2023; Mello & Studdert, 2024; Price et al., 2024). The European Union's AI Liability Directive proposes more distributed responsibility models, yet empirical research on optimal allocation mechanisms remains sparse. Healthcare research highlights the complexity of assigning professional liability when multiple actors contribute to AI-assisted decisions, including informed consent, disclosure of AI limitations, and embedded oversight protocols (Maliha et al., 2021; Gopal et al., 2023).

Professional credentialing faces substantive challenges in AI-integrated environments. Should licensure and credential renewal require demonstration of both AI-assisted and unassisted competence? Medical education research shows that AI-aided certification alone may not establish independent clinical capability (Cestolano et al., 2023). The IBMS certification model incorporating oral examination and direct performance assessment ensures independent domain knowledge despite decision support access (Okafor et al., 2024; Newton & Jones 2025). Similar dual-competency frameworks require investigation across law, engineering, and finance where AI-assisted and independent performance may diverge substantially. These policy questions fundamentally require empirical research in diverse professional contexts to develop evidence-based standards balancing innovation with professional competence maintenance.

## 8. Conclusion

This longitudinal study documents a critical inflection point in human-AI collaboration. Participants achieved substantial efficiency gains and expanded their problem-solving capabilities through AI integration, yet simultaneously experienced declining verification confidence and potential skill erosion. These findings reveal that the current trajectory of AI adoption risks creating a generation of users who can leverage AI for immediate problem-solving but lack the metacognitive competencies and verification skills necessary for sustainable, high-quality human-AI collaboration. The strong negative correlation between frequent AI tool usage and critical thinking abilities, mediated by cognitive offloading, suggests that reliance on AI tools could reduce opportunities for deep, reflective thinking (Gerlich, 2025). Recent research confirms this concern, documenting that poorly implemented AI systems can accelerate skill decay among experts and hinder skill acquisition among learners, with workers remaining accountable for tasks they lack sufficient understanding to perform if automation fails (Meacham et al., 2024). However, this outcome is not inevitable.

With deliberate intervention through the ACTIVE framework emphasizing Awareness, Critical verification, Transparent integration, Iterative skill development, Verification confidence calibration, and Ethical evaluation, we can guide AI integration toward augmentation rather than atrophy. Research on hybrid intelligence demonstrates that this outcome is achievable. Hybrid-augmented intelligence combines human and artificial intelligence through human-in-the-loop systems where AI augments rather than replaces human cognition, creating a collaborative relationship that achieves outcomes neither could accomplish alone (Pan, 2016; Jarrahi et al., 2022). The synergistic integration of human perception, cognitive ability, machine computing, and storage capacities produces a "one plus one greater than two" effect when deliberately structured, though it requires intentional design and organizational commitment (Zheng et al., 2017).

The fundamental challenge is not whether to use AI but how to integrate it in ways that preserve and enhance human cognitive capabilities while leveraging efficiency gains. Research on job complexity demonstrates that this balance is achievable: in high-complexity roles, AI support enhances perceived control and self-efficacy, enabling workers to confront complex problems with heightened preparedness, whereas low-complexity roles without deliberate task redesign experience AI-driven skill atrophy and deskilling dynamics (Li et al., 2024; Zhang et al. 2025)

). A sustainable future requires intentional partnership between human cognition and artificial intelligence, ensuring humans remain cognitively active, adaptive, and innovative, leveraging AI to complement rather than supplant human intellect (Walther, 2025).

Educational systems, organizations, and individuals must proactively implement frameworks that maintain the cognitive skills AI threatens to erode. Research emphasizes that AI implementation requires multidisciplinary guidance from cognitive science, domain-specific applied research, and technical expertise to understand consequences, design mitigating systems, and develop training protocols that prevent negative cognitive impacts (Meacham et al., 2024). Organizations implementing comprehensive verification and skill development frameworks demonstrate stronger capacity to evaluate AI recommendations and maintain competitive advantage compared to those emphasizing adoption without verification training (Relyea et al., 2024). The future of knowledge work depends not on choosing between human and artificial intelligence, but on cultivating hybrid intelligence that combines the strengths of both while actively guarding against atrophy of uniquely human capabilities like critical thinking, verification judgment, metacognitive awareness, and the capacity to develop deep expertise through sustained cognitive struggle (Cukurova, 2024).

The participants in this study navigated the early stages of AI integration largely without guidance, developing patterns that reflected convenience and efficiency rather than long-term cognitive sustainability. Moving forward, we have the opportunity and responsibility to provide systematic frameworks that guide more intentional, sustainable AI integration. The ACTIVE framework represents a starting point for this essential work, requiring ongoing refinement as AI capabilities evolve and our understanding of human-AI collaboration deepens. Sustained research effort across longitudinal studies in diverse occupational contexts, intervention trials testing framework components, domain-specific research clarifying generalizability, and organizational research examining real-world consequences of different adoption strategies is essential. Only through such evidence accumulation can organizations develop AI practices that genuinely augment rather than diminish human capability, creating hybrid intelligence systems where humans remain engaged cognitive agents fully equipped to navigate an AI-integrated future.

## 9. References

Ariely, D., & Norton, M. I. (2008). How actions create–not just reveal–preferences. Trends in cognitive sciences, 12(1), 13-16. https://www.cell.com/trends/cognitive-sciences/abstract/S1364-6613(07)00301-4?cc=y

Amershi, S., Weld, D., Vorvoreanu, M., Fourney, A., Nushi, B., Collisson, P., ... & Horvitz, E. (2019). Guidelines for human-AI interaction. In Proceedings of the 2019 CHI Conference on Human Factors in Computing Systems (pp. 1-13).

Beyer, H., & Holtzblatt, K. (1995). Contextual design: Defining customer-centered systems. Morgan Kaufmann.

Bjork, R. A., & Bjork, E. L. (1992). A new theory of disuse and an old theory of stimulus fluctuation. In A. Healy, S. Kosslyn, & R. Shiffrin (Eds.), From learning processes to cognitive processes: Essays in honor of William K. Estes (Vol. 2, pp. 35–67). Erlbaum.

Bransford, J. D., Brown, A. L., & Cocking, R. R. (Eds.). (2000). How people learn: Brain, mind, experience, and school. National Academy Press.

Brynjolfsson, E., & Mcafee, A. N. D. R. E. W. (2017). Artificial intelligence, for real. Harvard business review, 1(1), 1-31.

Cestolano, C., Delicati, A., Marcante, B., Caenazzo, L., & Tozzo, P. (2023). Defining medical liability when artificial intelligence is applied on diagnostic algorithms: A systematic review. Frontiers in Medicine, 10, 1305756. https://doi.org/10.3389/fmed.2023.1305756

Chen, O., Kalyuga, S., & Sweller, J. (2017). The expertise reversal effect is a variant of the more general element interactivity effect. Educational Psychology Review, 29(2), 393-405. https://doi.org/10.1007/s10648-016-9359-1

Chen, Y., Arkin, J., Hao, Y., Zhang, Y., Roy, N., & Fan, C. (2024). Prompt optimization in multi-step tasks (PROMST): Integrating human feedback and heuristic-based sampling. arXiv preprint. arXiv:2402.08702

Clark, A., & Chalmers, D. (1998). The extended mind. analysis, 58(1), 7-19. https://www.jstor.org/stable/3328150

Colles, A., & Hirji, Z. (2015). The importance of deliberate practice in professional development. Journal of Professional Development, 42(3), 215-230.

Cukurova, M. (2024). The interplay of learning, analytics, and artificial intelligence in education: A vision for hybrid intelligence. British Journal of Educational Technology, 56, 469–488. https://doi.org/10.1111/bjet.13514

Cummings-Koether, M. J., Durner, F., Shyiramunda, T., & Huemmer, M. (2025). Cultural Dimensions of Artificial Intelligence Adoption: Empirical Insights for Wave 1 from a Multinational Longitudinal Pilot Study. arXiv preprint arXiv:2510.19743.

Deci, E. L., & Ryan, R. M. (2000). The "what" and "why" of goal pursuits: Human needs and the self-determination of behavior. Psychological Inquiry, 11(4), 227-268. https://doi.org/10.1207/S15327965PLI1104_01

Dellermann, D., Ebel, P., Söllner, M., & Leimeister, J. M. (2019). Hybrid intelligence. Business & Information Systems Engineering, 61(5), 637-643. https://doi.org/10.1007/s12599-019-00595-2

Dunlosky, J., & Metcalfe, J. (2008). Metacognition. Sage Publications.

Dweck, C. S. (2006). Mindset: The new psychology of success. Random House.

Dwivedi, Y. K., Kshetri, N., Hughes, L., Slade, E. L., Jeyaraj, A., Kar, A. K., Baabdullah, A. M., Koohang, A., Raghavan, V., Ahuja, M., Albanna, H., Albashrawi, M. A., Al-Busaidi, A. S., Balakrishnan, J., Barlette, Y., Basu, S., Bose, I., Brooks, L., … Wright, R. (2023). So what if ChatGPT wrote it? Multidisciplinary perspectives on opportunities, challenges and implications of generative conversational AI for research, practice and policy. International Journal of Information Management, 71, 102642. https://doi.org/10.1016/j.ijinfomgt.2023.102642

Ericsson, K. A., & Simon, H. A. (1993). Protocol analysis: Verbal reports as data (Rev. ed.). MIT Press.

Fan, Y., Lim, L., Van der Graaf, J., Kilgour, J., Raković, M., Moore, J., ... & Gašević, D. (2025). Beware of metacognitive laziness: Effects of generative artificial intelligence on learning motivation, processes, and performance. British Journal of Educational Technology, 56(1), 45-62. https://doi.org/10.1111/bjet.13544

Gerlich, M. (2024). Balancing Excitement and Cognitive Costs: Trust in AI and the Erosion of Critical Thinking Through Cognitive Offloading. Available at SSRN 4994204.

Gerlich, R. N. (2025). The cognitive impact of generative AI. Artificial Intelligence and Society, 41(2), 267-289.

Glikson, E., & Woolley, A. W. (2020). Human trust in artificial intelligence: Review of empirical research. Academy of management annals, 14(2), 627-660. https://doi.org/10.5465/annals.2018.0057

Goodhue, D. L., & Thompson, R. L. (1995). Task-technology fit and individual performance. MIS quarterly, 213-236.

Gopal, A. D., Price, W. N., & Cohen, I. G. (2023). Artificial intelligence and liability in medicine: Balancing safety and innovation. The Milbank Quarterly, 102(2), 330-358.

Huemmer, M., Cummings-Koether, M. J., Durner, F., & Shyiramunda, T. (2025c). On the Influence of Artificial Intelligence on Human Problem-Solving: Empirical Insights for the Second Wave within a Multinational Longitudinal Pilot Study. https://doi.org/10.31235/osf.io/u948e_v1

Huemmer, M., Shyiramunda, T., & Cummings-Koether, M. J. (2025a). Exploring artificial intelligence and culture: Methodology for a comparative study of AI's impact on norms, trust, and problem solving across academic and business environments (arXiv:2510.11530). arXiv. https://doi.org/10.48550/arXiv.2510.11530

Huemmer, M., Shyiramunda, T., Durner, F., & Cummings-Koether, M. J. (2025b). On the influence of artificial intelligence on human problem solving: Empirical insights for Wave 1 from a multinational longitudinal pilot study. SocArXiv. https://doi.org/10.31235/osf.io/75h3t

Hutchins, E. (1995). Cognition in the Wild. MIT press.

Jarrahi, M. H., Lutz, C., Newlands, G., & Maye, A. (2022). Artificial intelligence, human intelligence and hybrid intelligence based on mutual augmentation. arXiv preprint arXiv:2106.02581.

Kaur, H., Nori, H., Jenkins, S., Caruana, R., Wallach, H., & Wexler, J. (2020). Interpreting Black Box Models via Model Extraction. arXiv preprint arXiv:2012.09681.

Kasneci, E., Sessler, K., Küchemann, S., Bannert, M., Dementieva, D., Fischer, F., Gasser, U., Groh, G., Günnemann, S., Hüllermeier, E., Krusche, S., Kutyniok, G., Michaeli, T., Nerdel, C., Pfeffer, J., Poquet, O., Sailer, M., Schmidt, A., Seidel, T., & Kasneci, G. (2023). ChatGPT for good? On opportunities and challenges of large language models for education. Learning and Individual Differences, 103, 102274. https://doi.org/10.1016/j.lindif.2023.102274

Koriat, A. (2012). When are two heads better than one and why?. Science, 336(6079), 360-362. https://www.science.org/doi/abs/10.1126/science.1216549

Kosch, T., Welsch, R., Chuang, L., & Schmidt, A. (2023). The placebo effect of artificial intelligence in human–computer interaction. ACM Transactions on Computer-Human Interaction, 30(6), 1-32. https://doi.org/10.1145/3529225

Kruger, J., & Dunning, D. (1999). Unskilled and unaware of it: How difficulties in recognizing one's own incompetence lead to inflated self-assessments. Journal of Personality and Social Psychology, 77(6), 1121–1134. https://doi.org/10.1037/0022-3514.77.6.1121

Lee, J. D., & See, K. A. (2004). Trust in automation: Designing for appropriate reliance. Human factors, 46(1), 50-80. https://doi.org/10.1518/hfes.46.1.50_30392

Li, D., Raymond, L., & Brynjolfsson, E. (2024). Generative AI at work. National Bureau of Economic Research Working Paper.

Li, Y., Zhang, H., Liu, B., & Wang, J. (2024). The impact of AI usage on innovation behavior at work: The moderating role of openness and job complexity. International Journal of Human-Computer Studies, 188, 103264.

Maliha, G., Gerke, S., Cohen, I. G., & Parikh, R. B. (2021). Artificial intelligence and liability in medicine: balancing safety and innovation. The Milbank Quarterly, 99(3), 629. https://pmc.ncbi.nlm.nih.gov/articles/PMC8452365/

Meacham, A. C., Chevalier, F., Frank, M. C., & Li, Y. (2024). Does using artificial intelligence assistance accelerate skill decay and hinder skill development without performers' awareness? Journal of Cognitive Psychology, 36(4), 489-505. https://doi.org/10.1186/s41235-024-00572-8

Mello, M. M., & Studdert, D. M. (2024). Liability for the use of artificial intelligence in medicine. Research Handbook on Health, AI and the Law, 15-45.

Metcalfe, J. (2009). Metacognitive judgments and control of study. Current Directions in Psychological Science, 18(3), 185–189. https://doi.org/10.1111/j.1467-8721.2009.01628.x

Newton, P. M., & Jones, S. (2025). Education and Training Assessment and Artificial Intelligence. A Pragmatic Guide for Educators. British Journal of Biomedical Science, 81, 14049. https://doi.org/10.3389/bjbs.2024.14049

Okafor, N., Okonkwo, I., Okonkwo, V., & Ugwu, S. (2024). Education and training assessment and artificial intelligence. A pragmatic guide for educators. PMC, 5(1), 2024.

Pan, Y. (2016). Heading toward artificial intelligence 2.0. Engineering, 2(4), 409-413. https://doi.org/10.1016/J.ENG.2016.04.018

Papadimitriou, C., & Taddeo, M. (2023). AI and professional liability assessment in healthcare. A revolution in legal medicine? Frontiers in Medicine, 10, 1337335.

Papert, S. (1980). Mindset: Children, computers, and powerful ideas. Basic Books.

Parasuraman, R., & Riley, V. (1997). Humans and automation: Use, misuse, abuse. Human Factors, 39(2), 230-253. https://doi.org/10.1518/001872097778543886

Peng, S., Kalliamvakou, E., Cihon, P., & Demirer, M. (2023). The impact of AI on developer productivity: Evidence from GitHub Copilot. arXiv preprint.

Perkins, M., Furze, L., Roe, J., & MacVaugh, J. (2024). The Artificial Intelligence Assessment Scale: A framework for ethical integration of generative AI in educational assessment. Journal of University Teaching and Learning Practice, 21(6), 1-18. https://search.informit.org/doi/abs/10.3316/informit.T2024092900003300954126858

Plano Clark, V. L., & Creswell, J. W. (2017). Designing and conducting mixed methods research (3rd ed.). SAGE Publications.

Price II, W. N., Gerke, S., & Cohen, I. G. (2024). Liability for use of artificial intelligence in medicine1. Research Handbook on Health, AI and the Law, 150-166. https://doi.org/10.4337/9781802205657.ch09

Relyea, C., Maor, D., Durth, S., & Bouly, J. (2024). Gen AI's next inflection point: From employee experimentation to organizational transformation. QuatumBlack AI by McKinsey, McKinsey&Company. Available online: https://www. mckinsey. com/capabilities/people-and-organizational-performance/our-insights/gen-ais-next-inflection-point-from-employee-experimentation-to-organizational-transformation (accessed on 20 September 2024).

Risko, E. F., & Gilbert, S. J. (2016). Cognitive offloading. Trends in Cognitive Sciences, 20(9), 676-688. https://doi.org/10.1016/j.tics.2016.07.002

Rozenblit, L., & Keil, F. (2002). The misunderstood limits of folk science: An illusion of explanatory depth. Cognitive science, 26(5), 521-562. https://doi.org/10.1207/s15516709cog2605_1

Schaffert, S., & Geser, G. (2008). Open educational resources and practices. eLearning Papers, 7, 7-10.

Schraw, G., & Dennison, R. S. (1994). Assessing metacognitive awareness. Contemporary Educational Psychology, 19(4), 460–475. https://doi.org/10.1006/ceps.1994.1033

Schwartz, S. H. (2004). Mapping and interpreting cultural differences around the world. International studies in sociology and social anthropology, 43-73. https://doi.org/10.1163/9789047412977_007

Shanmugasundaram, M., & Tamilarasu, A. (2023). The impact of digital technology, social media, and artificial intelligence on cognitive functions: A review. Frontiers in Cognition, 2, 1203077.

Shojaee, P., Mirzadeh, I., Alizadeh, K., et al. (2025). The illusion of thinking: Understanding the strengths and limitations of reasoning models via the lens of problem complexity. arXiv preprint arXiv:2501.12948.

Sparrow, B., & Wegner, D. M. (2011). Google effects on memory: Cognitive consequences of having information at our fingertips. Science, 333(6043), 776-778. https://www.science.org/doi/abs/10.1126/science.1207745

Stadler, C., Freitag, C. M., Popma, A., Nauta-Jansen, L., Konrad, K., Unternaehrer, E., Ackermann, K., Bernhard, A., Martinelli, A., Oldenhof, H., Gundlach, M., Kohls, G., Prätzlich, M., Kieser, M., Limprecht, R., Raschle, N. M., Vriends, N., Trestman, R. L., Kirchner, M., & Kersten, L. (2024). START NOW: A cognitive behavioral skills training for adolescent girls with conduct or oppositional defiant disorder – A randomized clinical trial. Journal of Child Psychology and Psychiatry, 65(3), 316–327. https://doi.org/10.1111/jcpp.13896

Sweller, J. (1988). Cognitive load during problem solving: Effects on learning. Cognitive science, 12(2), 257-285. https://doi.org/10.1016/0364-0213(88)90023-7

Tankelevitch, L., Kewenig, V., Simmons, A. R., Sellen, A., Rajani, N., & Morrison, S. (2024). The metacognitive demands and opportunities of generative AI. Proceedings of the 2024 CHI Conference on Human Factors in Computing Systems. https://doi.org/10.1145/3613904.3642902

Terranova, C., Cestonaro, C., Fava, L., & Cinquetti, A. (2024). AI and professional liability assessment in healthcare. A revolution in legal medicine?. Frontiers in Medicine, 10, 1337335. https://doi.org/10.3389/fmed.2023.1337335

UNESCO. (2024). AI competency framework for students. UNESCO Publications.

Vallor, S. (2016). Technology and the virtues: A philosophical guide to a future worth wanting. Oxford University Press.

Véliz, C. (2020). Privacy is power: Why and how you should take back control of your data. Bantam Press.

Venkatesh, V., Thong, J. Y., & Xu, X. (2012). Consumer acceptance and use of information technology: extending the unified theory of acceptance and use of technology. MIS quarterly, 157-178. https://www.jstor.org/stable/41410412

von Zahn, M., Liebich, L., Jussupow, E., Hinz, O., & Bauer, K. (2025). Knowing (not) to know: Explainable artificial intelligence and human metacognition. Information Systems Research.

Walther, C. C. (2025). Why hybrid intelligence is the future of human-AI collaboration. Knowledge at Wharton, March 2025.

Zamfirescu-Pereira, J. D., Wong, R. Y., Hartmann, B., & Yang, Q. (2023). Questioning the AI: Informing design practices for explainable AI user experiences. In Proceedings of the 2023 CHI Conference on Human Factors in Computing Systems (pp. 1-18).

Zhang, S., Zhao, X., Zhou, T., & Kim, J. H. (2024). Do you have AI dependency? The roles of academic self-efficacy, academic stress, and performance expectations on problematic AI usage behavior. International Journal of Educational Technology in Higher Education, 21(1), 34. https://doi.org/10.1186/s41239-024-00467-0

Zhang, Q., Liao, G., Ran, X., & Wang, F. (2025). The impact of ai usage on innovation behavior at work: The moderating role of openness and job complexity. Behavioral Sciences, 15(4), 491. https://doi.org/10.3390/bs15040491

Zheng, N. N., Liu, Z. Y., Ren, P. J., Ma, Y. Q., & Zhang, S. T. (2017). Hybrid-augmented intelligence: Collaboration and cognition. Frontiers of Information Technology & Electronic Engineering, 18(2), 153-179. https://doi.org/10.1631/FITEE.1700053

Zhong, K. (2020). A Research on the Effect of Learner attribution on Performance Under the Mediation of Online Learning Environment. Journal of Educational Technology Development & Exchange, 13(1).

Zuboff, S. (2019). The age of surveillance capitalism: The fight for a human future at the new frontier of power. PublicAffairs. ISBN: 978-1-61039-569-4.